%% file: main.tex
\documentclass[conference,compsoc]{IEEEtran}

\usepackage{support-caption}
\usepackage{subcaption}
\usepackage{amsmath,amsfonts,amssymb,amsthm,commath}
\usepackage{pifont}

\usepackage{microtype}
\usepackage{nicefrac}
\usepackage{graphicx,xcolor}
\usepackage{booktabs,multirow,multicol}
\PassOptionsToPackage{hyphens}{url}
\usepackage{hyperref}
\usepackage{cleveref}
\usepackage{complexity}
\usepackage{siunitx}
\usepackage{tikz,tikz-qtree}
\usepackage{bbding} 
\usepackage{lipsum} 
\usepackage[numbers,sort&compress]{natbib}
\usepackage{wrapfig}
\usepackage{enumitem}
\usepackage{textcomp}
\usepackage{xurl}
\usepackage{colortbl}
\usepackage{float}
\usepackage{xspace}
\usepackage{makecell}
\usepackage{wasysym}
\usepackage[official]{eurosym}
\definecolor{tabgray}{gray}{0.90}
\definecolor{tlegray}{gray}{0.5}
\usepackage{diagbox}

\usepackage{soul}
\usepackage[normalem]{ulem}
\renewcommand\sout{\bgroup\markoverwith
{\color{orange!90!black}{\rule[.5ex]{2pt}{1pt}}}\ULon}

\usepackage{pifont}
\newcommand{\cmark}{\ding{51}}%
\newcommand{\xmark}{\ding{55}}%

\usepackage{etoolbox}
\crefname{appendix}{}{}

\input{llymacro}

\newcount\Comments  
\newcount\arxiv 
\usepackage{color}
\definecolor{darkgreen}{rgb}{0,0.5,0}
\definecolor{darkblue}{rgb}{0,0,0.5}
\definecolor{purple}{rgb}{1,0,1}
\newcommand{\kibitz}[2]{\ifnum\Comments=0\textcolor{#1}{#2}\fi}

\usepackage[most]{tcolorbox}

\definecolor{boxcolor}{RGB}{230,240,250}

\newenvironment{takeaway}[1][]
  {
 \begin{tcolorbox}
 [%
    enhanced, 
    breakable,
    boxrule=0.5pt,
    arc=4pt,
    left=2pt,
    right=2pt,
    bottom=2pt,
    top=2pt,
    rounded corners
    ]{}
  \textbf{#1.}
  \small \itshape}
  {
\end{tcolorbox}
}

\ifnum\Comments=0
\fi

\ifnum\arxiv=1
\fi

\newcommand{\chang}[1]{{\color{red} Chang: #1}}

\input{macros}

\newcommand{\svmcls}{SVM\xspace}
\newcommand{\kmarginal}{K-Mgn\xspace}

\Crefname{definition}{Def.}{Defs.}
\Crefname{figure}{Fig.}{Figs.}

\begin{document}

\title{SoK: Privacy-Preserving Data Synthesis}

\author{Yuzheng Hu$^{*1}$\quad Fan Wu$^{*1}$\quad Qinbin Li$^2$\quad Yunhui Long$^1$\quad Gonzalo Munilla Garrido$^3$\\ Chang Ge$^4$\quad Bolin Ding$^5$\quad David Forsyth$^1$\quad Bo Li$^1$\quad Dawn Song$^2$\\
\small{$^1$University of Illinois Urbana-Champaign\quad 
$^2$UC Berkeley
}\\
\small{
$^3$Technische Universität München\quad
$^4$University of Minnesota\quad 
$^5$Alibaba Group
}
}


\maketitle


\setcounter{footnote}{1}
\footnotetext{
$^*$ indicates equal contribution.
Correspondence to: Yuzheng Hu $<$\href{mailto:yh46@illinois.edu}{yh46@illinois.edu}$>$ and Fan Wu   $<$\href{mailto:fanw6@illinois.edu}{fanw6@illinois.edu}$>$.
}

\input{contents/0_abstract}
\input{contents/1_intro}

\input{contents/2_preliminaries}

\input{contents/3_ppds}

\input{contents/4_deep}
\input{contents/6_evaluation}
\input{contents/7_future}
\newpage
{
\scriptsize
{\linespread{0.85}\selectfont\bibliography{ref}}
\bibliographystyle{ieeetr}
}

\onecolumn

\input{tables/tab_ref_2}
\appendix

\subsection{Method Selection in Real-World Scenarios}

Given the massive amount of literature, one natural question from \textit{practitioners} is how to efficiently locate the  
desired algorithm in real-world scenarios. Here we provide some  guidelines for method selection.

A real-world application of PPDS is mostly determined by the target \textit{data type} and \textit{task}. As such, these two factors can be used to
filter algorithms and obtain the candidate solutions as a first step for practitioners. 
In light of this observation, we present a scenario-oriented reference table (\Cref{tab:model_summary}), which classifies
the papers that we cover in~\Cref{sec:stat} and \ref{sec:deep} by data types and applicable tasks.  
Moreover, for each sublist of approaches, 
we identify the paper that has received the highest number of citations according to Google Scholar as ``Most-cited'' and the latest paper as ``Fresh''. 
We also identify the methodologies used in the most-cited and fresh papers. With Table \ref{tab:model_summary}, practitioners can easily locate the set of possible candidates as well as the more competent ones within  
based on their need in real-world scenarios. Moreover, practitioners can refer to Table \ref{tab:stat} and Table \ref{tab:deep} for the detailed properties of these candidates with the help of Table \ref{tab:model_summary}. 

\subsection{Meta-Review}

\subsubsection{Summary}
This is an SoK paper on the topic of privacy-preserving data synthesis (PPDS). The paper provides a comprehensive literature review, classifying existing data synthesis methods as either statistical methods or deep methods. The paper also proposes a ``master recipe'' that abstracts the components of a method and experimentally compares existing deep methods.

\subsubsection{Scientific Contributions}
\begin{itemize}
\item Other: SoK paper
\item Independent Confirmation of Important Results with Limited Prior Research
\end{itemize}

\subsubsection{Reasons for Acceptance}
\begin{enumerate}
\item The paper provides an overview of synthetic data generation and state of the art techniques for each type of data, which may be useful for newcomers to the area. The paper also describes open problems and future directions, which may be useful in shaping future research.

\item The paper evaluates six deep PPDS approaches on image synthesis using two standard datasets and identify DP-MERF as an all-purpose best approach. The paper also provides guidelines for practitioners who seek to use PPDS.

\item The reviewers ask that prior to publication, you update your main paper to motivate why there are no experiments on tabular data. For instance, you can indicate that the parts of the paper related to tabular data generation methods have already been studied and evaluated empirically in prior work, and provide the citation.
\end{enumerate}

\subsection{Response to the Meta-Review}

We accept the meta-review. We appreciate the valuable feedback from the reviewers and the shepherd.

\end{document}


\title{Full Experiment Results for\\ SoK: Certified Robustness for Deep Neural Networks}

\author{
}

\maketitle

\begin{abstract}
In this document, we provide full experiment results for the paper ``SoK: Certified Robustness for Deep Neural Networks'', including detailed experimental setup and full results.
\end{abstract}

    We evaluate and compare the robustness verification and training approaches by two sets of experiments:
    (1)~for major deterministic verification approaches, we compare their certified robustness over a diverse set of trained models of different scales;
    (2)~for major probabilistic verification approaches and their corresponding robust training approaches, we compare the best certified robustness they jointly achieve since the verification and training approaches are usually coupled.
    We do such separation mainly because in deterministic and probabilistic settings, the model inference methods are different and cannot be directly compared.
    
    The full details for probabilistic set of experiments have been provided in the paper.
    Here, we present the \emph{full experiment details} for the \emph{deterministic verification approach benchmark}.
    
    The open-sourced unified evaluation platform is available at \textit{\url{https://github.com/sokcertifiedrobustness/certified-robustness-benchmark}}.

    \section{Experimental Setup}
    
        \begin{table}[htbp]
            \caption{Deterministic verification approaches benchmarked in our evaluation.}
            \centering
            \resizebox{0.40\textwidth}{!}{
            \begin{tabular}{c|c|l}
                \toprule
                Category & Approach & Implementation \\
                \hline \hline
                MILP & \t{Bounded MILP}~\cite{tjeng2018evaluating} & Reimplementation \\
                \hline
                Branch and Bound & \t{AI$^2$}~\cite{gehr2018ai2} & From \cite{singh2019abstract} \\
                \hline
                 & \t{LP-Full}~\cite{weng2018towards,salman2019convex} & From \cite{boopathy2019cnn} \\
                  & \t{DeepPoly}~\cite{singh2019abstract} & From \cite{singh2019abstract} \\
                & \t{Fast-Lin}~\cite{weng2018towards} & From \cite{boopathy2019cnn} \\
                Linear & \t{CROWN}~\cite{zhang2018efficient} & From \cite{zhang2020towards} \\
                Relaxation & \t{CNN-Cert}~\cite{boopathy2019cnn}  & From \cite{boopathy2019cnn} \\
                 & \t{CROWN-IBP}~\cite{zhang2020towards} & From \cite{zhang2020towards} \\
                & \t{IBP}~\cite{gowal2019scalable} & Reimplementation \\
                & \t{WK}~\cite{wong2018provable,wong2018scaling} & From \cite{wong2018scaling} \\
                \hline
                \multirow{2}{*}{Hybrid}  & \t{k-ReLU}~\cite{singh2019beyond} & From \cite{singh2019abstract} \\
                 & \t{RefineZono}~\cite{singh2019beyond} & From \cite{singh2019abstract} \\
                \hline
                \multirow{2}{*}{SDP} & \t{SDPVerify}~\cite{raghunathan2018semidefinite} & Reimplementation \\
                & \t{LMIVerify}~\cite{fazlyab2019safety} & Reimplementation \\
                \hline
                \multirow{3}{*}{Lipschitz} & \t{Op-norm}~\cite{szegedy2013intriguing} & From \cite{zhang2019recurjac} \\
                & \t{Fast-Lip}~\cite{weng2018towards} & From \cite{zhang2019recurjac} \\
                & \t{RecurJac}~\cite{zhang2019recurjac} & From \cite{zhang2019recurjac} \\
                \bottomrule
            \end{tabular}
            }
            \label{table:exp-A-evaluated-verification-approaches}
        \end{table}
    
        \para{Evaluated Approaches.} 
        We present a thorough comparison of major \emph{deterministic verification approaches} as listed in \Cref{table:exp-A-evaluated-verification-approaches}.
        The list covers major branches of deterministic verification approaches.
        
        \para{Experiment Environment.} 
        For MNIST experiments, we run the evaluation on $24$-core Intel Xeon E5-2650 CPU running at \SI{2.20}{GHz} with single NVIDIA GeForce GTX 1080 Ti GPU.
        For CIFAR-10 experiments, we run the evaluation on $24$-core Intel Xeon Platinum 8259CL CPU running at \SI{2.50}{GHz} with single NVIDIA Tesla T4 GPU.
        The CIFAR-10 experiments use slightly faster running environment due to its larger scale.
        
        \para{Tool Implementation.}
        We implement our tool mainly in \texttt{PyTorch}~\cite{pytorch}.
        Some verification approaches use \texttt{keras}~\cite{keras} and \texttt{Tensorflow}~\cite{tensorflow}.
        For them, we implement a model translator from \texttt{PyTorch} directly to \texttt{keras} model, or to \texttt{onnx}~\cite{onnx} intermediate representation.
        The verification approaches reimplemented by ourselves use intrinsic \texttt{PyTorch} tensor computations, or \texttt{CVXPY}~\cite{cvxpy} + \texttt{Gurobi}~\cite{gurobi} solver when there is an optimization problem to solve.
        We test our implementation to ensure the correctness and high-efficiency.
        For verification approaches with hyperparameters, we use their default hyperparameters as described in the papers or original implementations.
    
        \para{Dataset.}
        We evaluate these verification approaches on image classification datasets MNIST~\cite{lecuyer2019certified} and CIFAR-10~\cite{krizhevsky2009learning}.
            
        \para{Models.}
        We include $7$ different neural network structures from the literature and adapt them for each dataset.
        Among them, $3$~(\sc{FCNNa} - \sc{FCNNc}) are fully-connected neural networks, and $4$~(\sc{CNNa} - \sc{CNNd}) are convolutional neural networks.
        They are all feed-forward ReLU neural networks.
        The number of neurons ranges from $50$~(\sc{FCNNa}) to about $200,000$~(\sc{CNNd}).
        For each neural network structure, we train $5$ sets of weights: 
        \begin{itemize}
            \item \texttt{clean}~(on both MNIST and CIFAR-10): regular training; 
            \item \texttt{adv1}~(on MNIST)/\texttt{adv2}~(on CIFAR-10): PGD adversarial training with $\epsilon = 0.1$ (on MNIST) or $\epsilon = 2/255$ (on CIFAR-10); 
            \item \texttt{adv3}~(on MNIST)/\texttt{adv8}~(on CIFAR-10): PGD adversarial training with $\epsilon = 0.3$ (on MNIST) or $\epsilon = 8/255$ (on CIFAR-10);
            \item \texttt{cadv1}~(on MNIST)/\texttt{cadv2}~(on CIFAR-10): \t{CROWN-IBP} training with $\epsilon = 0.1$ (on MNIST) or $\epsilon = 2/255$ (on CIFAR-10); 
            \item \texttt{cadv3}~(on MNIST)/\texttt{cadv8}~(on CIFAR-10): \t{CROWN-IBP} training with $\epsilon = 0.3$ (on MNIST) or $\epsilon = 8/255$ (on CIFAR-10).
        \end{itemize}
            
        \input{tables/exp-A-models-stats}
        \input{tables/exp-A-models-originalacc}
            
        Here, $\epsilon$ is the $\cL_\infty$ radius of the adversarial examples.
        The PGD adversarial training~\cite{madry2017towards} is a strong empirical defense.
        We use $40$-step PGD in PGD adversarial training.
        The \t{CROWN-IBP}~\cite{zhang2020towards} is a strong robust training approach which achieves state-of-the-art certified robustness under $(\cL_\infty, 8/255)$ adversary.
        We choose these training configurations to reflect $3$ common types of models on which  verification approaches are used: normal models; empirical defense models; and robustly trained models.
        All models are trained to reach their expected accuracy or robustness.
        The model structures and statistics of number of neurons is shown in \Cref{table:expA-models-stats}.
        The model accuracy on original test set, and empirical robust accuracy under the PGD attack, is shown in \Cref{table:expA-models-originalacc}.
            
        \para{Evaluation protocol.}
        We present the performance of verification approaches by their \emph{robust accuracy} and \emph{average certified robustness radius} with respect to $\cL_\infty$ radius $\epsilon$.
            
        \vspace{1em}
        \emph{Robust Accuracy:}
        When measuring robust accuracy,
        the chosen radius $\epsilon$ for verification corresponds to the radius trained for defense.
        The robust accuracy is defined as $\mathrm{robacc} := \dfrac{\text{\# samples verified to be robust}}{\text{\# number of all samples}}$.
        On each dataset, we uniformly sample $100$ test set samples as the fixed set for evaluation.
        We further limit the running time to \SI{60}{s} per instance and count timeout instances as ``non-robust''.
        
        We also report the robust accuracy under empirical attack~(PGD attack), which gives an upper bound of robust accuracy, so we can estimate the gap between certified robust accuracy and accuracy of existing attack.
        We remark that due to hyperparameter tuning, limited running time~(\SI{60}{s}), and different neural network models used in evaluation, the robust accuracy of some approaches differs much with the reported numbers, such as \t{k-ReLU}~\cite{singh2019beyond}, \t{RefineZono}~\cite{singh2018robustness}, \t{SDPVerify}~\cite{raghunathan2018semidefinite}, and \t{LMIVerify}~\cite{fazlyab2019safety}.
        
        \vspace{1em}
        \emph{Average Certified Robustness Radius:}
        We also evaluate the verification approaches by measuring their average certified robustness radius.
        The \emph{average certified robustness radius}~($\bar{r}$) stands for the average $\cL_\infty$ radius the verification approach can verify on the given subset of test set samples.
        We use the same uniformly sampled sets as in robust accuracy evaluation.
        To determine the best certified radius of each verification approach, we conduct a binary search process due to the monotonicity.
        Specifically, we do binary search on interval $[0,\,0.5]$ because the largest possible radius is $0.5$ for $[0,\,1]$ bounded input.
        If current radius $mid$ is verified to be robust, we update current best by $mid$ and let $l \gets mid$, if current radius $mid$ cannot be verified, we let $r \gets mid$, until we reach the precision $10^{-2}$ on MNIST or $10^{-3}$ on CIFAR-10, or time is up.
        For the evaluation of average certified robustness radius, since it involves multiple evaluations because of binary search, 
        we limit the running time to \SI{120}{s} per input and record the highest certified radius it has been verified before the time is used up.
        The average certified robustness radius is evaluated on the same subset of test set samples as used for robust accuracy evaluation.
        
        We also report the smallest radius of adversarial samples found by empirical attack~(PGD attack), which gives an upper bound of certified robustness radius for us to estimate the gap.
        
        \vspace{1em}
        Throughout the evaluation, we use $100$-step PGD attack with step size equals to $\epsilon / 50$ where $\epsilon$ is the constrained $\cL_\infty$ attack radius.
            
    \section{Full Results}
        
        We present the full results as below.
        Besides robust accuracy and average certified robustness radius, we also report the running time per instance for each approach on both tasks respectively.
        Note that for robust accuracy evaluation, the verification is run once per instance with time limit \SI{60}{s}; 
        while for average certified robustness radius evaluation, the verification will be run multiple times due to the binary search, with overall time limit \SI{120}{s}.
        
        We summarize the references of corresponding result tables as below.
        
        \begin{table}[H]
            \centering
            \begin{tabular}{cccc}
                \toprule
                Dataset & Evaluation Type & Results Table & Running Time Table \\
                \midrule
                \multirow{2}{*}{MNIST} & Robust Accuracy~($\mathrm{robacc}$) & \Cref{tab:exp-A-mnist-verify-robust-acc} & \Cref{tab:exp-A-mnist-verify-time} \\
                & Average certified robustness radius~($\bar r$) & \Cref{tab:exp-A-mnist-radius-robust-acc} & \Cref{tab:exp-A-mnist-radius-time} \\
                \hline
                \multirow{2}{*}{CIFAR-10} & Robust Accuracy~($\mathrm{robacc}$) & \Cref{tab:exp-A-cifar10-verify-robust-acc}\footnotemark & \Cref{tab:exp-A-cifar10-verify-time} \\
                & Average certified robustness radius~($\bar r$) & \Cref{tab:exp-A-cifar10-radius-robust-acc} & \Cref{tab:exp-A-cifar10-radius-time} \\
                \bottomrule
            \end{tabular}
            \centering
        \end{table}
        
        \footnotetext{The same table as appeared in the main paper.}
            
        \subsection*{Remarks.}
        Generally, the tendency on MNIST is the same as that on CIFAR-10.
        
        From these additional evaluations of average certified robustness radius, we can find that the average radius has better precision than robust accuracy.
        For example, on small models such as \sc{FNNa} and \sc{FNNb}, if measured by robust accuracy, the complete verification approaches and some linear relaxation-based approaches have almost the same precision~(see \cref{tab:exp-A-cifar10-verify-robust-acc}, the results of \t{Bounded MILP}, \t{AI$^2$}, \t{CROWN}, and \t{CNN-Cert}).
        However, if measured by average certified robustness radius, we can observe that complete verification approaches can certify much larger radius since they do not use any relaxations~(see \cref{tab:exp-A-cifar10-radius-robust-acc}, the result that \t{Bounded MILP} $\approx$ \t{AI$^2$} $>$ \t{CROWN} $\approx$ \t{CNN-Cert}).
        
        From the running time statistics, we further observe that a main cause of the failing cases~(i.e., $0$ robust accuracy or certified radius) for these approaches is the excessive verification time.
        In particular, since the evaluation of average certified robustness radius requires multiple times of verification for each instance, it suffers more severely from long verification time.

\bibliographystyle{IEEEtranSN}

{
\small
\bibliography{ref}
}
        
\newpage
        
        \input{tables/exp-A-mnist-verify-robust-acc}
        \input{tables/exp-A-mnist-verify-time}
        \input{tables/exp-A-mnist-radius-robust-acc}
        \input{tables/exp-A-mnist-radius-time}
        \input{tables/exp-A-cifar10-verify-robust-acc}
        \input{tables/exp-A-cifar10-verify-time}
        \input{tables/exp-A-cifar10-radius-robust-acc}
        \input{tables/exp-A-cifar10-radius-time}

%% file: llymacro.tex
\usepackage{dsfont}

\def\ddefloop#1{\ifx\ddefloop#1\else\ddef{#1}\expandafter\ddefloop\fi}
\def\ddef#1{\expandafter\def\csname bb#1\endcsname{\ensuremath{\mathbb{#1}}}}
\ddefloop ABCDEFGHIJKLMNOPQRSTUVWXYZ\ddefloop
\def\ddef#1{\expandafter\def\csname b#1\endcsname{\ensuremath{\mathbf{#1}}}}
\ddefloop ABCDEFGHIJKLMNOPQRSTUVWXYZ\ddefloop
\def\ddef#1{\expandafter\def\csname c#1\endcsname{\ensuremath{\mathcal{#1}}}}
\ddefloop ABCDEFGHIJKLMNOPQRSTUVWXYZ\ddefloop
\def\ddef#1{\expandafter\def\csname h#1\endcsname{\ensuremath{\hat{#1}}}}
\ddefloop abcdefghijklmnopqrsuvwxyz\ddefloop

\def\b1{\mathbf{1}}

\def\veps{\varepsilon}

\def\1{\mathds{1}}
\newcommand{\red}[1]{{\color{red} #1}}

\theoremstyle{definition}
\newtheorem{definition}{Definition}
\theoremstyle{remark}



%% file: macros.tex
\newcommand{\eg}[0]{\textit{e.g.}}
\newcommand{\ie}[0]{\textit{i.e.}}
\newcommand{\etal}[0]{\textit{et al.}\xspace}
\renewcommand{\sc}[1]{\textsc{#1}}

\usepackage{color}

\newcommand{\bo}[1]{\textcolor{magenta}{Bo: #1}}
\newcommand{\todo}[1]{\textcolor{blue}{TODO: #1}}
\newcommand{\yh}[1]{\textcolor{magenta}{#1}}

\newcommand{\mirnegg}[1]{\textcolor{blue}{(Yuzheng: #1)}}

\newcommand{\chulinr}[1]{\textcolor{brown}{#1}}
\newcommand{\fan}[1]{{\color{orange!90!black}(Fan: #1)}}

\newcommand{\eat}[1]{}

\newcommand{\dawn}[1]{\textcolor{red}{(Dawn: #1)}}
\newcommand{\varun}[1]{\textcolor{red}{(VC: #1)}}

%% file: contents/0_abstract.tex
\begin{abstract}


As the prevalence of data analysis grows, safeguarding data privacy has become a paramount concern. 
Consequently, there has been an upsurge in the development of mechanisms aimed at privacy-preserving data analyses.
However, these approaches are task-specific; designing algorithms for new tasks is a cumbersome process. As an alternative, one can create synthetic data that is (ideally) devoid of private information.
This paper focuses on privacy-preserving data synthesis (PPDS) by providing a comprehensive overview, analysis, and discussion of the field. 
Specifically, we put forth a {\em master recipe} that unifies  two prominent strands of research in PPDS: \textit{statistical methods} and \textit{deep learning (DL)-based methods}.
Under the master recipe, we further dissect the statistical methods into choices of modeling and representation, and investigate the DL-based methods by different generative modeling principles. To consolidate our findings, we provide comprehensive reference tables, distill key takeaways, and identify open problems in the existing literature.
In doing so, we aim to answer the following questions: What are the {\em design principles} behind different PPDS methods? How can we categorize these methods, and what are the \emph{advantages} and \emph{disadvantages} associated with each category? Can we provide \emph{guidelines} for method selection in different real-world scenarios? We proceed to benchmark several prominent DL-based methods on the task of private image synthesis and conclude that DP-MERF is an all-purpose approach. Finally, upon systematizing the work over the past decade, we identify future directions and call for actions from researchers.
\end{abstract}

%% file: contents/1_intro.tex
\vspace{-1mm}
\section{Introduction}
    \label{sec:intro}
\vspace{-1mm}

Recent years have witnessed an array of groundbreaking scientific achievements across various sectors, including but not limited to medical analyses~\cite{sidey2019machine,kononenko2001machine}, home automation and robots~\cite{chik2016review,siau2018building}, face recognition~\cite{kortli2020face,wang2021deep}, and malware detection systems~\cite{gavriluct2009malware,gibert2020rise}. 
These significant strides in practical applications can be attributed to the evolution of hardware technology, refinement of algorithms,
and most importantly---the availability of large-scale datasets for learning tasks or analytic workflows. 

While data serves as the critical fuel, the analysis of such large-scale datasets can often lead to privacy breaches. Examples range from exposing a driver's working hours and location through the data collected for training autonomous vehicles~\cite{glancy2012privacy,boeglin2015costs}, to the reconstruction of data used in training machine learning classifiers~\cite{fredrikson2015model,zhang2020secret}, and even the potential exposure of personal identities from medical images or records~\cite{fredrikson2014privacy,schwarz2019identification}.
The consequences are severe: the average cost of a healthcare data breach was of \$$9.2$M~\cite{data-breach-new}. 

In response to the escalating pace of data collection and analysis as well as the underlying privacy risks, regulatory bodies worldwide have enacted numerous legislations. Notably, the Health Insurance Portability and Accountability Act (HIPAA) of 1996 introduced the privacy rules for disclosing medical records~\cite{nosowsky2006health}, and the Federal Rules of Civil Procedure guide the disclosure of court records~\cite{tobias1988public}. These legislations have exerted considerable impact on various industries. For instance, non-compliance with HIPAA has resulted in substantial financial penalties, with fines reaching as high as \$$1.25$M~\cite{BannerHealthFine} and \$$6.9$M~\cite{PremeraBlueFine}.
Amidst the convergence of regulations, penalties, cybersecurity threats, and heightened user awareness~\cite{huth2020empirical}, the academic community has embarked on various research endeavors aimed at facilitating data analysis in a \textit{privacy-preserving} manner. These investigations encompass a range of paradigms, such as interactive and non-interactive query processing~\cite{brand2002microdata, blum2005practical}, data mining~\cite{vaidya2004privacy, aggarwal2008general, mendes2017privacy}, machine learning~\cite{ji2014differential,mohassel2017secureml,al2019privacy,boulemtafes2020review,liu2021machine}, among others.

However, creating privacy-preserving solutions tailored to each specific data analysis task is a formidable challenge that necessitates extensive customization.
An alternative involves synthesizing a dataset {devoid of private information} that can be used for {\em any} downstream task, enabling ``\textit{exploratory, open-ended, and iterative}''~\cite{hay2010enabling} analyses. In contrast to the aforementioned paradigms, synthetic data permits the execution of a variety of downstream tasks without requiring modifications to the original algorithms to meet privacy constraints. This offers flexibility in terms of future task options, whilst mitigating concerns over privacy leakage.  

In this paper, we focus on such privacy-preserving data synthesis (PPDS) approaches. 
 It's important to note that the uninformed application of privacy mechanisms to the original data could result in significant utility loss~\cite{dankar2012application}. To this end, PPDS mechanisms require delicate design choices to maintain the privacy-utility trade-off.

\noindent\textbf{Motivation.}\quad 
Research of PPDS in the past decade has demonstrated a shift of focus compared to earlier works due to the following \textit{emerging} challenges:
\begin{itemize}[leftmargin=*]
\item \textbf{High-Dimensional and Unstructured Data.} \quad Classical PPDS approaches designed for tabular data~\cite{mohammed2011differentially, blum2013learning} encounter obstacles in coping with the rising dimensionality of data and the inherent complexity involved in handling unstructured data such as images and text.
    
\item \textbf{Advanced Attacks.} \quad
Traditional anonymization-based PPDS approaches which adopt either the syntactic privacy~\cite{clifton2013syntactic} or less rigorous privacy principles are known to suffer from various privacy  attacks~\cite{narayanan2006break, wong2007minimality, narayanan2008robust, ohm2009broken, fiore2020privacy}.

\end{itemize}

\begin{figure*}
    \centering
    \includegraphics[width=.85\textwidth]{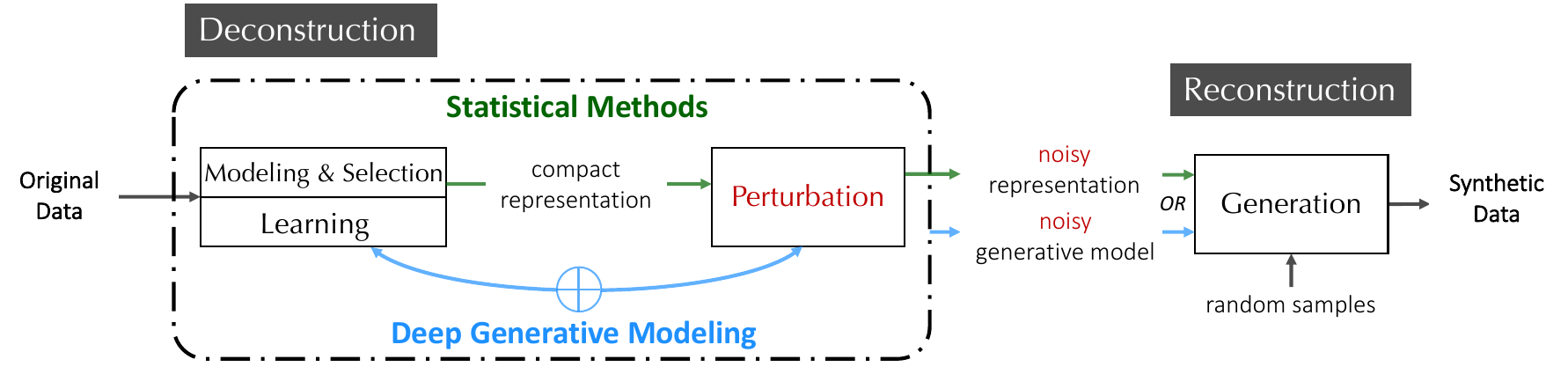}
    \caption{\small \textbf{The master recipe.} 
    In the deconstruction stage, statistical methods obtain a compact representation through \textit{modeling \& selection}, which is then subject to perturbations.
    On the other hand, DL-based methods typically integrate  \textit{learning} with perturbation. The resulting noisy representation or generative model is then used to generate synthetic data in the reconstruction stage.
        }
    \label{fig:framework}
\end{figure*}

To alleviate concerns associated with the data-related challenges, statistical models (\eg, hidden Markov model (HMM)~\cite{eddy2004hidden} and Bayesian network~\cite{friedman1997bayesian}) and tools (\eg, copula~\cite{nelsen2007introduction}) have been used to facilitate the modeling and deconstruction of high-dimensional structured data, while deep generative models~\cite{taylor2021deep,goodfellow2014generative,kingma2013auto} arise as powerful tools to handle unstructured data. To remediate concerns on the privacy front, research is focused on developing rigorous privacy notions (e.g., differential privacy (DP)~\cite{DP_original}) with corresponding privacy mechanisms, as well as their adaptations and applications in different scenarios. With abundant work exploring the intersection of these two lines of research at hand, we believe it is time for us to take a step back and systematically review the established toolkits and the lessons learned. 

\noindent {\bf{Standing Point and Contributions.}}\quad Our work joins a small family of surveys (see ``Related Work'' below) that study PPDS. While we are not the first to review this topic, we position our work as \textbf{the first to systematize the research effort over the past decade which tackles the emerging challenges using the most up-to-date toolkits}. Specifically, we make the following contributions:

\begin{itemize}[leftmargin=*]
\item We present a comprehensive overview of PPDS (\Cref{sec:prelim}), considering different privacy notions and mechanisms, input data types as well as utility evaluation protocols. 

\item We distinguish two primary research directions in PPDS: statistical methods and deep learning (DL)-based methods. Furthermore, we present a master recipe (\Cref{fig:framework}) that unifies these approaches, encompassing the algorithms falling within each category.
\item Under the master recipe, 
we dissect the statistical methods (\Cref{sec:stat}) into choices of modeling and representation, and investigate the DL-based methods (\Cref{sec:deep}) by different generative modeling principles. To consolidate our findings, we provide comprehensive reference tables, distill key takeaways, and identify open problems in the existing literature.
\item We benchmark the DL-based methods for private image synthesis (\Cref{sec:eval}), and identify DP-MERF~\cite{harder2021dp} as the best practice. We further note that the pessimistic conclusion on synthetic tabular data by Stadler \etal~\cite{stadler2021synthetic} does not fully extend to the image domain.
\item We delineate future directions for researchers (\Cref{sec:future}).
\end{itemize}

\noindent\textbf{Related Work.} \quad
Prior surveys related to PPDS can be classified into two categories: (i) privacy-preserving data publishing (PPDP)~\cite{zhou2008brief, wu2010survey, chen2009privacy, fung2010privacy, fiore2020privacy} which focuses on the anonymization of structured data, and (ii) privacy-preserving deep generative models~\cite{fan2020survey, cai2021generative}. Apart from these works, we also identify a few benchmarks~\cite{bowen2019comparative, tao2022benchmarking} for some representative algorithms on private tabular data synthesis, 
as well as a general introduction to synthetic data~\cite{jordon2022synthetic}.

\noindent\textbf{Scope.} \quad In terms of the \textit{timeline}, we collect papers after 2010 which allows us to concentrate on DP, and also distance ourselves from a prior survey~\cite{fung2010privacy}. Our {\em ontology} is as follows: \Cref{sec:stat} primarily concerns the intersection of database and privacy, whereas \Cref{sec:deep} concerns the intersection of DL and privacy. In terms of the \textit{target}, we exclude works that \textit{only} facilitate learning tasks or analytic workflows~\cite{liu2008privacy, wang2013differential, li2015matrix, mckenna2018optimizing}, or those that publish aggregated or summary statistics~\cite{barak2007privacy, qardaji2014priview} rather than generating synthetic data.


%% file: contents/2_preliminaries.tex
\vspace{-2mm}
\section{Preliminaries}
\label{sec:prelim}
\vspace{-2mm}

Given an input dataset $D$, we aim to design a PPDS algorithm $P$, such that $P(D) \rightarrow D'$ with $D'$ being the synthesized dataset which is both $\textit{private}$ and $\textit{utilitarian}$. 
Formally,  $P$ satisfies a specific {\textbf{privacy} definition} (\Cref{subsec:privacy-require}); and  $D'$ attains high \textbf{utility} (\Cref{subsec:types/utilities}). 

\vspace{-2mm}
\subsection{Privacy Notions and Mechanisms}
\label{subsec:privacy-require}
\vspace{-2mm}

There are numerous definitions of privacy spanning different applications and time-periods~\cite{westin1968privacy, warren1989right, wu2013defining, cohen2020towards}. In the context of PPDS, we adopt the perspective of F. T. Wu~\cite{wu2013defining}, which posits that privacy remains intact if no violation, in the form of \textit{re-identification}, occurs.
We deal with a broad sense of re-identification beyond human identities~\cite{porter2008identified,el2011systematic,malin2004not,rocher2019estimating} and consider the exposure of any sensitive information within data
as re-identification. 
This could range from unveiling connections among users in social networks~\cite{zheleva2007preserving}, to revealing sensitive facial attributes, such as the eyes, nose, or mouth, in facial image data~\cite{newton2005preserving}.

We focus on \textit{differential privacy} (DP), which provides mathematical guarantees against re-identification. 
Privacy definitions such as {syntactic anonymity}~\cite{clifton2013syntactic} (\eg, $k$-anonymity~\cite{sweeney2002k} and its variants) and less rigorous privacy principles such as {obfuscation}~\cite{brunton2015obfuscation} were widely adopted in PPDS~\cite{fung2010privacy, fiore2020privacy}. However, recently, there is a clear tendency to prefer DP~\cite{ji2016graph, nist2018, nist2020} because of (i) its ability to quantify privacy leakage, and (ii) the stronger protection it offers~\cite{cohen2022attacks}. 

\noindent\textbf{Differential Privacy (DP).}\quad
Along with a variety of benign properties including composability~\cite{dwork2009differential}, immune to post-processing~\cite{dwork2014algorithmic} and future-proof~\cite{garfinkel2021implementing}, DP has become the gold standard of privacy. DP quantifies and limits the amount of knowledge one could learn by observing the algorithm's output. Formally:
\begin{definition}[$(\varepsilon, \delta)$-DP~\citep{dwork2014algorithmic}]
\label{def:eps-delta-dp}
Let $\mathcal{X}$ be the space of the dataset, 
a randomized algorithm $\mathcal{M}$ with domain $\mathbb{N}^{|\mathcal{X}|}$ is $(\varepsilon,\delta)$-differentially private iff for all $\mathcal{S} \subseteq \text{Range}({\mathcal{M})}$ 
and for any neighboring datasets $D$ and $D'$ differing by one record:
$
    \Pr[\mathcal{M}(D) \in \mathcal{S}] \leq e^\varepsilon\Pr[\mathcal{M}(D')\in \mathcal{S}] + \delta.
$
\end{definition}
The standard definition assumes a centralized database with trusted server and {independence} between
participants~\cite{clifton2013syntactic}. 
R\'{e}nyi DP~(RDP)~\cite{mironov2017renyi} is a variant of DP that uses R\'{e}nyi divergence~\cite{renyi1961measures} to measure the distance between two probability distributions.
Other variants include  
edge-DP and node-DP~\cite{hay2009accurate} for graphs (see~\Cref{subsec:graph}). 

\noindent \textbf{Privacy Mechanisms.} For algorithms with numeric output, DP and its variants are typically enforced by adding noise (Laplace and Gaussian mechanism~\cite{dwork2014algorithmic}). For non-numeric purpose, conventional wisdom is to perform probabilistic selection based on a score function (exponential mechanism~\cite{mcsherry2007mechanism}).
For algorithms that involve iterative optimization of a parametrized model, DP-SGD~\cite{Abadi2016DeepLW} adds Gaussian noise to the clipped gradient.
Moments Accountant~\cite{Abadi2016DeepLW} and RDP accountant~\cite{mironov2017renyi} track the privacy cost throughout iterative optimization.




\vspace{-2mm}
\subsection{Data Types and Utility} \label{subsec:types/utilities}
\vspace{-2mm}
PPDS has engaged extensively with both \textit{structured} (S) and \textit{unstructured} (U) data, each necessitating distinct approaches. Statistical methods have been devised primarily for structured data, whereas DL-based methods are more suitable for unstructured data. 
We focus on the following data types:
(S1) \textit{tabular data}, which arrange elements in columns and rows with each column corresponding to an attribute and each row representing a record; (S2) \textit{trajectory data}, which comprise sequences of locations such as (latitude, longitude) pairs and reflect the movement of individual over time; (S3) \textit{graph data}, which consist of nodes connected by edges where the nodes denote certain entities and the edges represent their  interconnections; (U1) \textit{image data}, which are comprised of pixels. 


\textbf{Data utility} is characterized by a metric that depends on two primary factors: (i) the type of data, and (ii) the nature of the data analysis conducted. We take tabular data as a representative example. When the analysis task involves the construction of a histogram with the data, utility can be assessed by the similarity (e.g., $\ell_\infty$-norm) w.r.t. the outcome of queries, such as random linear queries~\cite{xu2017dppro}, range queries~\cite{hardt2012simple,li2014differentially,zhang2021differentially}, parity queries~\cite{gaboardi2014dual}, and aggregation queries~\cite{fan2020relational}. Conversely, when the analysis task entails training a machine learning model with the data, utility can be evaluated by the model’s performance on downstream tasks, for instance, classification accuracy~\cite{torkzadehmahani2019dp,chen2020gs,harder2021dp}. Due to space constraints, we defer a detailed summary of utility metrics to our GitHub repository\footnote{\url{https://sok-ppds.github.io/data_utility_and_fidelity.html}}.

Another concept that bears a close relationship to data utility is \textit{fidelity}. It refers to the extent to which inherent data properties (\eg, statistical or structural) are preserved, and is frequently applied in the literature to evaluate synthetic data. In assessing fidelity, similarity functions are often used for structured data, while pre-trained neural networks are employed for unstructured data.
We will delve further into the correlation between utility and fidelity in~\Cref{sec:eval}.
\input{tables/tab_sec3}

%% file: tables/tab_sec3.tex
\begin{table*}
\caption{\small \textbf{Statistical methods for PPDS.}
Generality: \CIRCLE=general, \Circle=specific. 
Scalability: \CIRCLE=good, \RIGHTcircle=moderate, \Circle=poor. 
\cmark=with, \xmark=without.
$/$=not applied. 
}
\label{tab:stat}
\renewcommand{\arraystretch}{1.2}
\setlength{\tabcolsep}{3pt}
\resizebox{\textwidth}{!}{%
\begin{tabular}{llllllclcccl}
\toprule
\multicolumn{2}{c}{\textbf{Data}} & \multicolumn{2}{c}{\textbf{Methodology}} & \multicolumn{3}{c}{\textbf{Privacy}} & \multicolumn{2}{c}{\textbf{Properties}} & \multicolumn{2}{c}{\textbf{Evaluation}} & \multicolumn{1}{c}{\textbf{References}} \\ \cmidrule(lr){1-2}\cmidrule(lr){3-4} \cmidrule(lr){5-7} \cmidrule(lr){8-9} \cmidrule(lr){10-11} 
\makecell[l]{Data\\ Type} & \makecell[l]{Data\\ Characteristics} & \makecell[l]{Approach/\\Model} & Tool& \makecell[l]{Privacy \\ Guarantee } &   \makecell[l]{Privacy \\ Mechanism } &
\makecell[l]{Privacy \\ Test } &
\makecell[l]{Utility\\ Metrics } & \makecell[l]{Utility \\ Guarantee} & Generality &  Scalability &  \\ \cline{1-12}
\multirow{7}{*}{\makecell[l]{Tabular\\ Data}} & \multirow{3}{*}{\makecell[l]{categorical\\ attribute}} & \multirow{2}{*}{marginal-based} & Bayesian network & DP& LM, (EM)&\cmark & S1-F1, F5, A1 & \xmark & \CIRCLE & \RIGHTcircle & \citep{zhang2017privbayes, bindschaedler2017plausible}  \\ \cline{4-12}
 &  &  & Markov random field & DP& LM / GM & \xmark& S1-F1, F2, A1 & \xmark & \CIRCLE & \RIGHTcircle & \citep{chen2015differentially, mckenna2021winning, zhang2021privsyn, cai2021data}  \\ 
 \cline{3-12}
 &  & query-based &/  & DP& EM, (LM) & \xmark & S1-F1, F2 & \cmark & \Circle & \Circle & \makecell[l]{\cite{hardt2012simple, gaboardi2014dual}\\\cite{vietri2020new, aydore2021differentially}} \\ 
 \cline{2-12}
 & \multirow{4}{*}{\makecell[l]{numerical\\ attribute}} & \multirow{2}{*}{copula-based} & Gaussian copula & DP& LM & \xmark &  S1-F1, F2 & \xmark & \CIRCLE & \Circle & \citep{li2014differentially, asghar2019differentially} \\ \cline{4-12}
 &  &  & vine copula & DP& LM & \cmark & \makecell[l]{S1-F1, F3, F6,\\A1, A2} & \xmark & \CIRCLE & \Circle & \citep{gambs2021growing} \\ \cline{3-12}
 &  & \multirow{2}{*}{projection-based} & random projection & attribute-DP& GM & \xmark &  S1-F2, F4, A1 & \cmark & \Circle & \CIRCLE & \citep{kenthapadi2012privacy, xu2017dppro}  \\ \cline{4-12}
 &  &  & others & DP& LM & \cmark & S1-F3, A1, A2, A3 & \xmark & \CIRCLE & \RIGHTcircle & \citep{li2011compressive, jiang2013differential, chanyaswad2019ron}  \\ 
 
 \cline{1-12}
\multirow{2}{*}{\makecell[l]{Trajectory\\Data}} & \makecell[l]{sequential\\ data} & tree-based &/  & DP& LM & \cmark & S2-F1$\sim$F7 & \xmark & \CIRCLE & \Circle & \citep{chen2012differentially, chen2012ngrams, zhang2016privtree, he2015dpt, wang2017protecting} \\ \cline{2-12}
 &  / & distribution-based & / & DP& LM & \cmark & S2-F1, F2, F4$\sim$F7 & \xmark & \CIRCLE & \RIGHTcircle &  \citep{mir2013dp, roy2016practical, gursoy2018differentially, gursoy2018utility, gursoy2020utility} \\ \cline{1-12}
\multirow{5}{*}{\makecell[l]{Graph\\ Data}} & \multirow{4}{*}{/} & SKG & / &edge-DP& LM & \xmark &  S3-F1, F4, F8, F9 & \xmark & \CIRCLE & \Circle & \citep{mir2012differentially}\\  \cline{3 - 12}
 &  & dK-Graph & / & edge-DP& LM & \xmark &  \makecell[l]{S3-F1$\sim$F4, F6,\\F8, F9, A1, A2} & \xmark & \CIRCLE & \RIGHTcircle & \citep{sala2011sharing, wang2013preserving}  \\  \cline{3 - 12}
 &  & ERGM & / & edge-DP& LM & \xmark & S3-F7, F8 & \xmark & \CIRCLE & \Circle & \citep{lu2014exponential}  \\  \cline{3 - 12}
 &  & HRG & / & edge-DP& EM, LM & \xmark &  S3-F1, F3, F9 & \xmark & \CIRCLE & \RIGHTcircle & \citep{xiao2014differentially}  \\   \cline{2 - 12}
 & \makecell[l]{attributed\\ graph} & AGM & / & edge-DP& LM & \xmark & S3-F1, F4, F5, F8 & \xmark & \CIRCLE & \RIGHTcircle & \citep{jorgensen2016publishing, chen2020publishing} \\
 \bottomrule
\end{tabular}%
}
\end{table*}

%% file: contents/3_ppds.tex
\vspace{-2mm}
 \section{PPDS via Statistical Methods}
 \label{sec:stat}
 \vspace{-2mm}
 
 \textbf{Overview.} \quad
In this section, we introduce the statistical methods for PPDS.
Unlike DL-based methods, which are capable of performing end-to-end synthesis, statistical methods are typically tailored to specific data types. Consequently, it is generally impractical to expect a single algorithm to function effectively across various scenarios without adjustments. To that end, we structure this section based on different data types. Our analysis encompasses three categories of structured data (S1, S2, S3) as outlined in 
\Crefrange{subsec:tabular}{subsec:graph}, 
respectively, with Table~\ref{tab:stat} providing a comprehensive overview of the literature from multiple perspectives. We exclude unstructured data from our discussion due to the lack of sufficient references.


\textbf{Master Recipe.} \quad 
Despite the variations in tools and implementation details, the majority of works discussed in this section can be encapsulated within a unified framework (refer to Figure~\ref{fig:framework}), which consists of two primary components: deconstruction and reconstruction. We refer to this as the \textit{master recipe} in the subsequent discussion. In the context of statistical methods, the deconstruction component involves modeling the original data and identifying a compact representation, potentially safeguarded by privacy measures. The subsequent noisy representation, derived through the application of Laplace or Gaussian mechanisms, is then passed to the reconstruction component for synthetic data generation. It is worth noting that the deconstruction step can effectively tackle the challenge posed by high-dimensionality by reducing the amount of noise required to ensure DP.

Most naturally, we unfold our discussion by examining how various approaches can be interpreted as instantiations of the master recipe, with a focus on the modeling of the original data as well as the choice of compact representation. 
 \subsection{PPDS for Tabular Data}
 \label{subsec:tabular}

We begin with tabular data, a prevalent data type in practical applications.
Due to its highly-structured nature, numerous approaches target tabular data from distinct perspectives while adhering to the same master recipe. 

  \subsubsection{\textbf{Paper Summaries}} \label{subsubsec:stat_cat} \quad
  We identify four approaches 
  of DP 
  tabular data synthesis based on marginal, copula, projection and query. At a high level, marginal-based and copula-based approach model the rows of tabular data as  \textit{realizations of a joint distribution}; projection-based approach models tabular data holistically as a \textit{matrix}; query-based approach belongs to a line of {theoretical} research and does not perform any modeling.

  \noindent \textbf{Marginal-Based Approach.} \quad
  The low-order marginals are the go-to choice of statistics for tabular data. They are adept at capturing the low-dimensional structure of real-world distributions and have low sensitivity, making them an ideal candidate for the compact representation in~\Cref{fig:framework}.
  In line with the master recipe, algorithms fall under the marginal-based approach roughly follow the same routine: select marginals, {measure} with noise, and generate synthetic data from the joint distribution estimated by the noisy marginals. To facilitate marginal selection and/or generation, the procedure is implemented within graphical models such as Bayesian network~\cite{friedman1997bayesian} or Markov random field~\cite{metzler2005markov}.

          \underline{\textit{Bayesian Network.}} Zhang~\etal~\cite{zhang2017privbayes} 
          propose PrivBayes, which simultaneously selects marginals and determines the topology of the Bayesian network by iteratively maximizing information gain through the exponential mechanism. Synthetic data can subsequently be sampled from the joint distribution, which is computed based on the Bayesian network topology. 
          Bindschaedler~\etal~\cite{bindschaedler2017plausible}    
          adopt a similar procedure as PrivBayes, albeit applied to only a random subset of attributes for \textit{partial} data synthesis. Additionally, the authors propose a privacy test for the release of synthesized records, which leads to the guarantee of plausible deniability~\cite{bindschaedler2016synthesizing}.

          \underline{\textit{Markov Random Field.}} 
          The Markov random field (MRF) is capable of capturing more complex correlations within the data compared to a Bayesian network.
          Chen~\etal~\cite{chen2015differentially} construct a graph that explore pairwise dependence between attributes and apply the junction tree algorithm to obtain the MRF, from which the noisy marginals are generated and the synthetic data are sampled.
        The probabilistic graphical-model (PGM) based estimation~\cite{mckenna2019graphical} does not fulfill PPDS per se, but is rather a reconstruction algorithm that can effectively learn a 
        MRF that best matches the noisy marginals
        via {mirror descent}~\cite{beck2003mirror}. 
            As a matter of fact, PGM serves as a crucial building-block in the \textit{state-of-the-art} mechanisms of tabular data synthesis: it gives rise to the top-ranked teams in the NIST 2018~\cite{nist2018} and 2020~\cite{nist2020} contests~\cite{mckenna2021winning, cai2021data} when coupled with  sophisticated techniques of marginal selection. More recently,  Graduate Update Method (GUM)~\cite{zhang2021privsyn} arises as an alternative  reconstruction algorithm
            to handle graphical models with dense interactions.

    {The marginal-based approach explicitly models the correlations between different attributes within graphical models, which
    makes the synthetic data capable of handling a wide range of tasks with satisfactory utility.} 

  \noindent \textbf{Copula-Based Approach.}\quad A copula, as defined within the realms of probability theory and statistics, refers to the joint cumulative distribution function which describes the \textit{dependence} among random variables. Once integrated with univariate marginal distributions, it is capable of determining the joint distribution (see Sklar's Theorem~\cite{sklar1959fonctions}),
 which forms the foundation for sampling synthetic data. Therefore, a combination of the univariate marginal distributions and a copula would serve as the compact representation in~\Cref{fig:framework}, and the key question is to determine the copula (the marginal distributions can be estimated from the histograms of the original data). We will discuss two classes of copula respectively, namely, Gaussian copula{~\cite{bouye2000copulas}} and
  vine copula{~\cite{bedford2001probability}}. 
  
         
    \underline{\textit{Gaussian Copula.}}       Gaussian copula enjoys a closed-form and can be uniquely determined by the correlation matrix. 
    DPCopula~\cite{li2014differentially} 
    considers two methods: maximum likelihood and Kendall's $\tau$  coefficient~\cite{demarta2005t} to estimate the correlation matrix, perturb with noise and derive the Gaussian copula thereof.
          To overcome the limited applicability of copula on discrete attributes,
          Asghar~\etal~\cite{asghar2019differentially} propose
          an additional step of pre-processing which can be applied to both categorical and numerical attributes before the estimation.
          
        \underline{\textit{Vine Copula.}}  Despite the simplicity and convenience of the Gaussian copula, it has been noted that the tail dependencies might not be fully captured by this model, and can lead to a significant 
        loss of information \cite{mackenzie2014formula}. 
          Therefore,  COPULA-SHIRLEY~\cite{gambs2021growing} adopts \textit{vine copula}, which decomposes the multivariate function into multiple bivariate copulas, and applies a top-down greedy 
          algorithm from~\cite{dissmann2013selecting} for copula selection.
          While being less computational efficient, the customizable framework as well as the ability of vine copula in modeling complex distributions lead to improved applicability and utility.
          
          {Similar to the marginal-based approach, copulas can adapt to the inherent 
          structure of the data, enabling algorithms to attain relatively high utility.} 
          But copula-based approach comes with the downside of high complexity---quadratic in the number of attributes in the case of Gaussian copula selection and even higher for vine copula.
          This renders them less  favorable compared to the marginal-based approach.

  \noindent \textbf{Projection-Based Approach.} \quad 
  Under the projection-based approach, tabular data is plainly modeled as a matrix, and the projections onto low-dimensional subspace naturally serve as the compact representation in~\Cref{fig:framework}. 
  Furthermore, the selection of these projections typically leverages certain geometric properties inherent to the high-dimensional space.

  
  \underline{\textit{Random Projection.}} There are a few works~\cite{kenthapadi2012privacy, xu2017dppro} that \textit{partially} align with the master recipe due to the absence of a reconstruction process. Specifically, the synthetic data are obtained via projecting the original data by row using a Gaussian random matrix followed by some perturbation. The pairwise $L_2$-distances between individuals are preserved by the Johnson-Lindenstrauss lemma~\cite{johnson1984extensions}, making the synthetic data suitable for tasks such as clustering and nearest neighbors which only rely on such information. 
  
  \underline{\textit{Others.}} Apart from random projection, other algorithms that fall under this category are consistent with the master recipe. Li~\etal~\cite{li2011compressive} obtain a sparse representation of the original data using a Bernoulli random matrix, and reconstruct the synthetic data from the perturbed representation via compressive sensing~\cite{candes2006robust}.
  Jiang~\etal~\cite{jiang2013differential} adopt the  
  noisy top principal components to construct the synthetic data, which are computed based on the perturbed first and second moments.
  Chanyaswad~\etal~\cite{chanyaswad2019ron} observe that the \textit{orthonormal} projection of high-dimensional data are nearly Gaussian~\cite{meckes2012projections}, and apply a Gaussian generative model on top of the noisy projection to produce the synthetic data. 
  
{On the positive side, algorithms based on projection are easy to implement and {usually} come with some utility \textit{guarantee}: they aim at preserving specific properties 
of the original data, and the corresponding utility bounds arise from the similarity in statistical measures. }
 {However, they usually 
  fail to capture the underlying structure of the original data, mainly due to the coarse selection of the compact representation which leads to a significant loss of information. As a consequence, the synthetic data generated from these algorithms are incapable of accommodating generic usage.}
  
  \noindent \textbf{Query-Based Approach.} \quad The query-based approach~\cite{hardt2012simple, gaboardi2014dual, vietri2020new, aydore2021differentially} is built upon a series of prior work that lay down the theoretical foundation of differentially private query processing~\cite{kasiviswanathan2011can, blum2013learning}, with the goal of privately releasing a synthetic dataset that can accurately answer an \textit{assigned} workload of linear queries instead of making it generically useful. While queries are a generalization of marginals and can be similarly interpreted as the compact representation in~\Cref{fig:framework}, there is no explicit modeling of the original data, and the reconstruction step deviates from the master recipe.
  As such, we will not discuss these algorithms in detail.

\begin{takeaway}[Takeaways]
    \textbf{Trade-off between specificity and generality}---Algorithms tailored to specific data usage, such as random projection, typically offer provable utility guarantees but lack versatility across different tasks. Conversely, marginal-based and copula-based approaches, while being heuristic in utility, excel at capturing attribute correlations. Moreover, their joint-probability modeling enables the generation of an \textit{infinite} amount of data.

\end{takeaway}

\begin{takeaway}[Open Problems]
\begin{itemize}[leftmargin=*]
    \item \textbf{Fine-grained structures.}\quad 
State-of-the-art algorithms for differentially private tabular data synthesis primarily rely on low-order marginals, which may not adequately capture the high-order correlations between attributes (\eg, denial constraints~\cite{chu2013discovering} and (conditional) functional dependencies~\cite{yao2008mining, li2013effective}). Integrating these fine-grained structures into the design of algorithms holds the potential to enhance data utility, a prominent example being Kamino~\cite{ge2021kamino}.
    
    
    \item \textbf{Mixed attributes.}\quad 
    In real-world scenarios, datasets commonly consist of a combination of categorical and numerical attributes. However, most existing algorithms are specialized for handling only one type of attribute (marginal-based approaches for categorical attributes and copula-based approaches for numerical attributes). Consequently, there is a pressing need for a thoughtful pre-processing design to generate tabular data with mixed attributes.
    
\end{itemize}

\end{takeaway}

\subsection{PPDS for Trajectory Data}
\label{subsec:trajectory}
While trajectory data typically take a tabular form, their ubiquitous collection in real-world scenarios~\cite{vines2017exploring, latif2020leveraging} has fostered the development of privacy-preserving trajectory data synthesis as a distinct subject of interest, which attracts substantial research efforts from the academic community. 
Compared to tabular data, trajectory data 1) embody \textit{temporal dependencies}, 
akin to general sequential data; 2)  incorporate \textit{key features} such as start and end points, trajectory length, and so forth. 
As such, in additional to discussing different approaches under the master recipe, we will also underscore how these features are integrated into the algorithmic design.
\subsubsection{Paper Summaries}  \quad
\label{subsubsec:traj_cat}
Following~\cite{fiore2020privacy}, we discuss two approaches of DP trajectory data synthesis: the tree-based approach and the distribution-based approach.
At a high level, these two approaches diverge in their modeling of trajectory data. The former primarily addresses the temporal dependency, viewing trajectory data broadly as a sequence. In contrast, the latter interprets trajectories as the combination of certain key features.

\noindent\textbf{Tree-Based Approach.} \quad
The tree-based approach attempts to create a noisy \textit{prefix tree} based on the original data, from which the synthetic trajectory data will be derived. The prefix tree---a hierarchical structure that groups trajectories with the same prefix into the same branch, can therefore be interpreted as the compact representation in~\Cref{fig:framework} and encodes the {temporal dependency} of the trajectory data.  

\underline{\textit{General Sequential Data Synthesis.}} A few works on differentially private trajectory data synthesis are designed for general sequential data. 
 Chen~\etal~\cite{chen2012differentially} 
 apply multiple levels of location generalization to the original prefix tree and add Laplacian noise to each generalized node, whose expansion terminates when the noisy count fails a certain threshold or a predefined tree height is reached. 
A subsequent work~\cite{chen2012ngrams} constructs the noisy prefix tree based on the $n$-gram model, which makes use of the Markov independence assumption 
to estimate the probability of a current location based on the previous $(n-1)$ ones. Compared to~\cite{chen2012differentially} whose majority of nodes have small counts due to the uniqueness of patterns, the $n$-gram model captures more temporal dependency and is therefore more robust to noise.
Building upon the Markov model, 
Zhang~\etal~\cite{zhang2016privtree} employ a \emph{prediction suffix tree}~\cite{begleiter2004prediction} which bears resemblance to the prefix tree explored in~\cite{chen2012differentially}.
Nonetheless, their algorithm brings several improvements to the table: it permits adaptability in the tree's height and incorporates a more sophisticated strategy for node expansion, which considers the diversity of potential future child nodes.
Both of these adaptations contribute to enhanced utility.

\underline{\textit{Fine-grained Trajectory Synthesis.}} 
The aforementioned algorithms 
only work with coarse trajectories on small location domain.
To produce fine-grained trajectories over real-world geographical span, follow-up works seek to leverage additional properties or features of trajectory data beyond temporal dependency.
He~\etal~\cite{he2015dpt} propose DPT, which takes into account the \textit{non-uniformity} of real trajectories: the algorithm
discretizes the trajectories using multiple location sampling 
resolutions
to capture individual movements at different speeds, and maintains a prefix tree for each resolution. 
Wang and Sinnott~\cite{wang2017protecting} establish the noisy prefix tree based on a private reference system. This construction calibrates original trajectories against a selection of anchor points, adeptly reflecting both overarching mobility patterns and detailed attributes such as turning and congestion points.

{The tree-based approach compactly summarizes the temporal dependency between consecutive locations and works more generally for sequential data.}
{However, algorithms that primarily rely on the tree structure usually fail to preserve useful properties  
such as the trajectory length.
Furthermore, they typically exhibit limited scalability. For instance, the discretization of a vast location domain elevates both the size of the prefix tree and the necessary noise~\cite{chen2012differentially}. Correspondingly, algorithms that depend on the Markov model~\cite{chen2012ngrams, zhang2016privtree} exhibit increasing inefficiencies as the trajectory length expands.



\noindent\textbf{Distribution-Based Approach.} \quad
Distribution-based approach first derives the probability distributions 
associated with key attributes of the original data, samples from the perturbed distributions, 
and generates synthetic trajectories thereof. The probability distributions can be viewed as the compact representation in~\Cref{fig:framework}.

Mir~\etal~\cite{mir2013dp} put forth DP-WHERE, an approach that estimates various distributions of an attribute set, capable of comprehensively determining human movement trajectories. Roy~\etal~\cite{roy2016practical}, on the other hand, consider the interactions between different attributes by grouping strongly correlated attributes into non-disjoint sets and constructing a corresponding distribution for each set.
More recently, Gursoy et al. (2018) introduce DP-Star, an approach that deconstructs trajectory data into four principal features: space, start and end points, transition probabilities, and route length. Each feature is associated with a probability distribution that is perturbed with noise. Synthetic trajectories are then reconstructed from the features sampled  from the noisy distributions. 
Two subsequent studies, AdaTrace~\cite{gursoy2018utility} and OptaTrace~\cite{gursoy2020utility}, adopt similar pipelines as~\cite{gursoy2018differentially} while offering respective enhancements in terms of privacy and utility.
Specifically, AdaTrace enforces deterministic constraints on the generated trajectories while maintaining DP, thereby bolstering the algorithm's resilience against common threats in trajectory data (\eg, Bayesian inference). OptaTrace employs Bayesian optimization to identify the optimal allocation of the privacy budget across the four features, aiming to minimize the output error relative to a specific metric (\eg, query error).
 

{Distribution-based approach leverages domain-specific features, equipping algorithms under this category to generate trajectories with high fidelity.}
{One minor disadvantage of the distribution-based approach is that it can \textit{occasionally} be outperformed by the tree-based approach~\cite{chen2012ngrams, he2015dpt} under stringent privacy settings (\eg, $\varepsilon = 0.1$). 
This occurs because the total privacy budget is divided into multiple shares, and the required noise escalates super-linearly with the decreasing budget.
As such, each component incurs substantial error when $\varepsilon$ is small.}

\begin{takeaway}[Takeaways]
    \textbf{The power of domain information}---Through a comparative analysis of the tree-based and distribution-based approaches, as well as the two subcategories within the tree-based approach, we observe that the inclusion of fine-grained properties (\eg, non-uniformity) or features (\eg, anchor points) consistently enhances utility and applicability. This serves as compelling evidence for the influential role of domain information in algorithmic design.
    
\end{takeaway}
\begin{takeaway}[Open Problems]
\begin{itemize}[leftmargin=*]
    \item \textbf{Expanding formats.}\quad Real-world trajectory data often encompasses supplementary columns that extend beyond spatiotemporal information, such as user attributes (\eg, occupation). The increasing prevalence of drones and airborne sensors will push data synthesis from 2D planes to 3D spaces. These diverse data formats present researchers with novel challenges.
    
    
    
    \item \textbf{Task-centric utility metrics.} \quad Currently, the evaluation of synthetic trajectories is \textit{statistical} in nature: the utility metrics primarily focus on fidelity but fail to capture the overall usefulness of the generated data.
    We envision the establishment of task-centric utility metrics as a vital step towards achieving equitable quality assessment across different algorithms.

\end{itemize}

\end{takeaway}

\subsection{PPDS for Graph Data}
\label{subsec:graph}
 Graphs surface in numerous real-life scenarios such as social networks and communication data~\cite{ji2016graph}. 
 Two variants of DP are proposed for this data structure: \emph{edge-DP} and \emph{node-DP}~\cite{hay2009accurate}, and we focus on the former since node-DP is scarce in the literature.
 \subsubsection{Paper Summaries} \quad
 \label{subsubsec:graph_cat}
 In line with the master recipe, 
 most algorithms for DP graph data synthesis share the same pipeline: extract graph statistics, which serve as the compact representation in~\Cref{fig:framework}, and generate the synthetic graph based on the perturbed statistics.   
 Additionally, the graph statistics are usually computed within a specific graph model, which reflects the \textit{belief} of the underlying generation process. As such, we organize the discussion based on different graph models and omit works that do not adopt any~\cite{proserpio2012workflow, chen2014correlated, mulle2015privacy, nguyen2015differentially}. 
 
 
 \noindent \textbf{Stochastic Kronecker Graph Model.} \quad  
 The Stochastic Kronecker Graph Model (SKG)~\cite{leskovec2007scalable} is generated through the Kronecker power of an adjacent matrix and reduces graph synthesis to parameter estimation.
 Mir and Wright~\cite{mir2012differentially} incorporate the Laplace mechanism into the moment-based estimation of SKG~\cite{gleich2012moment} to ensure edge-DP.   
 {While being simple to implement, }{SKG cannot accurately capture the structural properties of real graph data since the generation process is determined by a \textit{single} parameter.}
 
 \noindent \textbf{DK-Graph Model.} \quad 
 The dK-graph model~\cite{mahadevan2006systematic}  condenses the graph into the degree distribution of connected components of size $K$ (dK-series). Notably, this can also serve to reconstruct the graph when $K=2$, making the dK-$2$-series a primary research focus.
 Sala~\etal~\cite{sala2011sharing} 
observe that the global sensitivity of the dK-$2$-series is large and propose a partitioning-based algorithm, which divides the dK-$2$-series into different sub-series and adds \textit{non-uniform} noise based on their local sensitivity~\cite{dwork2006calibrating}.
However, it is highlighted in~\cite{nissim2007smooth} that adding noise based on local sensitivity might inadvertently expose private information.
As such, a subsequent work by Wang and Wu~\cite{wang2013preserving} calibrates the noise based on the smooth sensitivity~\cite{nissim2007smooth}, thereby providing a rigorous edge-DP guarantee.
{However, despite these improvements, the
privacy budget is still required to be unreasonably large
(\eg, $\varepsilon \ge 100$).}
 
 \noindent \textbf{Exponential Random Graph Model.} \quad The Exponential Random Graph Model (ERGM)~\cite{snijders2006new} is characterized by a vector of sufficient statistics, capable of encapsulating a broad spectrum of structural and attribute information.  
 Analogous to KGM, graph synthesis can be reformulated as a problem of parameter estimation in ERGM.
 Lu and Miklau~\cite{lu2014exponential} first privately compute the alternating graph statistics~\cite{snijders2006new}, and then estimate the model parameters using Bayesian inference~\cite{caimo2011bayesian}. 
 {However, ERGM can only work with a few thousand nodes and fail to scale to moderately-large graphs due to the heavy computational cost of Bayesian inference. }
 
 \noindent \textbf{Hierarchical Random Graph.} \quad 
 The hierarchical random graph (HRG)~\cite{clauset2008hierarchical} model represents a graph with a dendrogram which reflects its hierarchical structure, and some connection probabilities associated with the internal nodes of the dendrogram.
Xiao~\etal~\cite{xiao2014differentially} privately sample a dendrogram via Markov-Chain Monte Carlo (MCMC)~\cite{metropolis1953equation}, and propose
a thresholding strategy to calculate the noisy connection probabilities.
{Compared to other works, the HRG model achieves the smallest global sensitivity and therefore requires the least amount of noise to ensure edge-DP. This showcases the power of encoding structural information in the form of edge probabilities.}

 
{The models discussed so far are designed for unlabeled graphs and focus on the \textit{graph structure} in isolation. In contrast, real-world graph data such as social networks attach attributes to each vertex (e.g. affiliations on Facebook and Twitter), and their correlations are reflected by the graph structure in the form of \textit{homophily}---the tendency for nodes with similar attributes to form connections.} This observation guides us to the subsequent graph model. 
 
 \noindent \textbf{Attributed Graph Model.} \quad The attributed graph model (AGM)~\cite{pfeiffer2014attributed} attaches binary attributes (e.g. smoker v.s non-smoker) to nodes and performs conditional sampling of the graph structure based on the correlations between attributes.  
 AGM requires an generative model to capture the structural properties of the input graph, and contains three sets of parameters that characterize the distribution of attributes and edges as well as the correlations between attributes and edges. 
 Jorgensen~\etal~\cite{jorgensen2016publishing} propose a structural model called TriCycLe, which more effectively captures the clustering property of the input graph compared to its predecessors~\cite{chung2002average, pfeiffer2012fast}. The authors integrate TriCycLe into AGM and generate the synthetic graph based on privately-estimated parameters.
 A follow-up work by Chen~\etal~\cite{chen2020publishing} proposes the Community-Preserving Attributed Graph Model (C-AGM), which preserves the community structure while retaining the advantages of the AGM model.

\begin{takeaway}[Takeaways]
\textbf{Absence of a universal solution}---A universally applicable differentially private graph synthesis algorithm has yet to be found and might not even exist. Each method demonstrates distinctive trade-offs related to privacy, utility, and the preservation of specific graph properties. As a result, the selection of an appropriate algorithm is contingent on the demands and constraints inherent to each application. 
\end{takeaway}
\begin{takeaway}[Open Problems]
    \begin{itemize}[leftmargin=*]
    \item \textbf{Leveraging prior knowledge.} \quad 
The selection of graph models often lacks principled justification (except for AGM, which explicitly encodes attribute information). To enhance model selection in diverse scenarios, a combination of comprehensive understanding of each model and prior knowledge regarding the underlying generation process can prove invaluable.

    
    
    
    \item \textbf{Node-DP.} \quad
    Existing research has primarily centered around Edge-DP, which safeguards the relationship between users. However, node-DP offers a more robust notion of adjacency, where neighboring graphs differ in both nodes and associated edges. This introduces additional challenges in terms of selecting and fitting graph models, as it significantly impacts the graph statistics. Addressing this challenge remains an intriguing open problem for future investigation.
    
\end{itemize}

\end{takeaway}

%% file: contents/4_deep.tex
\input{tables/tab_GAN_utility_privacy}
\vspace{-2mm}
\section{PPDS via DL-Based Methods}
\label{sec:deep}
\vspace{-2mm}



\looseness=-1

Deep generative models are renowned for their capability and flexibility in handling high-dimensional data and generating diverse, high-fidelity samples~\cite{kingma2013auto,goodfellow2014generative,ho2020denoising,song2021scorebased}. We refer to these strategies as Deep Learning (DL)-based methods, given the prevalent use of modern neural networks (NNs) as their primary computational framework.
Although these models afford a degree of empirical privacy protection~\cite{lu2019empirical,li2019evaluating}, the substantial capacity of NNs can give rise to the ``memorization'' phenomenon~\cite{arpit2017closer}, thus rendering them vulnerable to membership inference attacks~\cite{hayes2017logan,hilprecht2019monte,xu2022mace}. Consequently, it is crucial to develop deep generative models with formal privacy guarantees.


\textbf{Master Recipe.}\quad
We discuss how privacy-preserving deep generative modeling aligns with the master recipe.
In the deconstruction stage, a generative model is learned from the training data. In the reconstruction stage, synthetic data is generated by feeding random samples to the generative model. DP is achieved by incorporating perturbations into the learning procedure.

The instantiations of the master recipe for DL-based methods markedly deviate from those for statistical methods across various dimensions, owing to the inherent differences in their working mechanisms.
Both the \textit{modeling} and the \textit{representation selection} in statistical methods are contingent upon the data type. 
In contrast,  
the \textit{generative modeling principle} plays a central role in DL-based methods.
Once the principle is determined and the model architecture is chosen (\eg, deep convolutional GANs for image data), the {representation selection} is automatically accomplished within the learning process.
Therefore, privacy enforcement in DL-based methods is marginally associated with the data type, but closely related to the generative modeling principle. 
Additionally, the iterative nature of the optimization algorithms employed in DL makes DP-SGD~\cite{Abadi2016DeepLW} the default privacy mechanism for DL-based methods. 


\looseness=-1
\textbf{Overview.}\quad
We organize the section based on different principles 
of generative modeling, including the auto-encoding mechanism, minimax training, optimal transport, feature alignment, and stochastic simulation (\Crefrange{subsec:deep-ae}{subsec:deep-ss}). Notably, some algorithms leverage multiple principles to perform PPDS. In this case, we categorize them based on the primary principle driving their design.
\Cref{tab:deep} provides a comprehensive overview of the literature.



\input{contents/4.2_vae}
\input{contents/4.1_adversarial}

\input{contents/4.5_optimal}
\input{contents/4.3_features}
\input{contents/4.4_stochastic}

%% file: tables/tab_GAN_utility_privacy.tex
\begin{table*}[htb]
\caption{\small \textbf{DL-based methods for PPDS.}
Generality: \CIRCLE=general, \Circle=specific. 
Scalability: \CIRCLE=good, \RIGHTcircle=moderate, \Circle=poor. 
Privacy Test: \cmark=with, \xmark=without. 
}\label{tab:deep}
\renewcommand{\arraystretch}{1.1}
\resizebox{\textwidth}{!}{%
\begin{tabular}{lllllcclcl}
\toprule
\multirow{2}{*}{\makecell[l]{\textbf{Generative}\\ \textbf{Modeling}\\\textbf{Principles}}} &
\multicolumn{2}{c}{\textbf{Privacy}} & \multicolumn{2}{c}{\textbf{Instantiations}} & \multicolumn{2}{c}{\textbf{Properties}} & \multicolumn{2}{c}{\textbf{Evaluation}} &  \multirow{2}{*}{\textbf{References}} \\
\cmidrule(lr){2-3} \cmidrule(lr){4-5} \cmidrule(lr){6-7}  \cmidrule(lr){8-9} 
 & \makecell[l]{Privacy\\ Mechanism} & \makecell[l]{Privacy\\ Accountant} & Data Type & Model Type & \multicolumn{1}{l}{{Generality}} & \multicolumn{1}{l}{Scalability} &   \multicolumn{1}{l}{{\makecell[l]{Utility \\Evaluation}}} & \multicolumn{1}{l}{\makecell[l]{Privacy \\ Test}} \\ \midrule

\multirow{4}{*}{\makecell[l]{Auto-Encoding\\Mechanism}} & {\makecell[l]{DP-SGD\\(+DP-$k$means)}} & MA & tabular, images & ensemble+VAE & \CIRCLE & \RIGHTcircle & S2-F1, U1-A1 & \xmark & \cite{acs2018differentially,chen2018differentially} \\
\cline{2-10}
 & {\makecell[l]{DP-SGD+DP-EM}} & MA & tabular & ensemble+AE & \CIRCLE & \Circle & S1-F1, S1-A1 & \xmark & \cite{abay2018privacy}\\
\cline{2-10} 
 & \multirow{2}{*}{DP-SGD} & MA & time series & GAN+AE & \CIRCLE & \Circle & S1-A1, S1-A3 & \xmark & \cite{lee2020generating} \\
 \cline{3-10}
 && RDP & \makecell[l]{tabular,\\network traffic} & GAN+AE & \CIRCLE & \Circle & \makecell[l]{S1-F1, S1-A1} & \xmark & \cite{tantipongpipat2021differentially,torfi2022differentially,fan2021dpnet}\\
 \hline
  \multirow{9}{*}{\makecell[l]{Minimax\\Training}} & \multirow{5}{*}{\makecell[l]{DP-SGD}} & \multirow{4}{*}{MA} & \multirow{3}{*}{\makecell[l]{tabular,\\network traffic}} & \makecell[l]{GAN, WGAN,\\ improved WGAN} & \CIRCLE & \Circle & \makecell[l]{S1-F1, F3, F4, \\ A1, A2, A3} 	& \cmark & \cite{lu2017poster,fan2020relational,liu2019ppgan,xie2018differentially,frigerio2019differentially,fan2021dpnet} \\ 
\cline{5-10} 
 &  &  &  & AC-GAN & \CIRCLE &  \Circle &S1-A1	&\xmark& \cite{beaulieu2019privacy} \\ \cline{5-10} 
 &  &  &  & CTGAN & \Circle & \RIGHTcircle & S1-A1	&	\xmark	& \cite{astolfi2021generating,wang2020part} \\ 
 \cline{4-10} 
 &  &  & images & \makecell[l]{WGAN,\\improved WGAN} & \CIRCLE & \Circle  &U1-F1, F4, A1	&	\cmark	& \cite{xie2018differentially,zhang2018differentially} \\ 
 \cline{3-10} 
 &  & \multirow{1}{*}{RDP} & \multirow{1}{*}{images} & 
  CGAN & \CIRCLE & \RIGHTcircle   &U1-A1, A3	&	\xmark	& \cite{torkzadehmahani2019dp} \\   
 \cline{2-10}
 & \makecell[l]{Partial Gradient\\Sanitization}  & RDP & images & improved WGAN & \CIRCLE & \RIGHTcircle  & U1-F1, F2, A1	&	\xmark	& \cite{chen2020gs} \\ 
 \cline{2-10} 
 & \multirow{3}{*}{\makecell[l]{Private Knowledge\\Aggregation and \\Transfer}} &
 MA & tabular & teacher-student+WGAN & \CIRCLE & \Circle  & S1-A1, A2	&	\xmark	& \cite{yoon2018pategan} \\ \cline{3-10} 
 &  & \multirow{2}{*}{RDP} & tabular, images & teacher-student+DCGAN & \CIRCLE & \RIGHTcircle  &
 \makecell[l]{ S1-A1, A2, \\ U1-F1, F2, A1}	&	\xmark	& \cite{long2021gpate} \\ \cline{4-10} 
 &  &  & images & teacher-student+DCGAN & \CIRCLE & \CIRCLE  & U1-A1	&	\xmark	&\cite{wang2021datalens} \\ \hline
  \makecell[l]{Optimal\\Transport} & \makecell[l]{Partial Gradient\\Sanitization} & RDP & images & generator & \Circle & \CIRCLE & U1-F2, U1-A1 & \xmark & \cite{cao2021dpsinkhorn}\\
 \hline
 \makecell[l]{Feature\\Alignment}
 & GM & RDP & tabular, images & generator & \CIRCLE & \RIGHTcircle & \makecell[l]{S1-A1, U1-A1,\\U1-F2, U1-F3}  & \xmark & \cite{harder2021dp,vinaroz2022hermite,harder2022differentially} \\
 \hline
 \multirow{2}{*}{\makecell[l]{Stochastic\\Simulation}} & RR & N/A & images & SGM & \Circle & \CIRCLE  &U1-F1, F2, A1	&	\xmark	& \cite{chen2022dpgen} \\
 \cline{2-10}
 & DP-SGD & RDP & images & Score SDE & \Circle & \CIRCLE & U1-F2, U1-A1 & \xmark & \cite{dockhorn2022dpdm} \\
 \bottomrule
\end{tabular}%
}
\end{table*}

%% file: contents/4.2_vae.tex
\subsection{PPDS Based on Auto-Encoding Mechanism}
\label{subsec:deep-ae}

\subsubsection{Preliminaries} \quad
Auto-encoders (AEs)~\cite{rumelhart1985learning} are designed to distill the data into a low-dimensional representation known as the \textit{latent space}, which
captures the core features of the data. The generation of new samples can be accomplished by sampling points from this latent space and subsequently decoding them back into the original data space. The latent space plays a central role in the design of AE-based approaches; lacking structure can lead to the creation of unrealistic samples.

\subsubsection{Paper Summaries} \quad
We review how different works impose structure on the latent space.

\noindent\textbf{Gaussian Latent Space.}\quad
Variational Autoencoders (VAEs)~\cite{kingma2013auto} impose a Gaussian prior on the latent space and optimize the parameters to minimize a combined loss, comprising the {reconstruction error} which measures the dissimilarity between the input and reconstructed data, as well as the KL divergence that quantifies the disparity between the prior and posterior distributions of the latent variables. This ensures a continuous and smooth structure in the latent space. One can apply DP-SGD to the end-to-end training of the VAE to achieve DP~\cite{acs2018differentially,chen2018differentially}.

\noindent\textbf{Gaussian Mixture Latent Space.}\quad
Imposing a Gaussian prior on the latent space of VAEs can restrict their expressiveness. To address this issue, several studies model the latent space with more flexible distributions.
DP-SYN~\cite{abay2018privacy} trains an AE using DP-SGD, and then models its latent space as a Gaussian Mixture Model (GMM) via DP Expectation Maximization (DP-EM)~\cite{park2017dp}.

\noindent\textbf{GAN-Learned Latent Space.}\quad
Rather than imposing a prior on the latent distribution, an automated approach uses GANs (see~\Cref{subsec:deep-at}) to learn the latent distribution of the trained AE~\cite{lee2020generating,tantipongpipat2021differentially,torfi2022differentially,fan2021dpnet}.
One can sample from the GAN to obtain latent representations of the AE, and then decode them to generate the synthetic data.
DP-SGD is applied to the training of both the AE and the GAN. 
This line of research aims to harness the strengths of both GANs and AEs to effectively generate \textit{discrete} data, such as tabular data with categorical attribute.
GANs facilitate the sampling from the AE's latent representation, and AEs permit the learning of discrete data, a task that cannot be accomplished solely by GANs.



\begin{takeaway}[Takeaways]
Different types of latent spaces entail unique trade-offs.
Gaussian latent spaces excel in structural simplicity and smoothness, but struggle in modeling complex data.
Enhancing the expressiveness and flexibility of the latent space comes at the cost of increased parametrization, leading to additional privacy cost. Therefore, the selection of latent space hinges on the specific use case. 

\end{takeaway}
\begin{takeaway}[Open Problems]
AEs, designed for efficient data representation, implicitly capture important features of the data distribution in the latent space. 
This characteristic is particularly suitable for PPDS, in which 
the utility of the synthesized data often outweighs their fidelity (\ie, similarity in data distribution).
To amplify this advantage, one promising approach could be to explicitly enhance the utility of the latent space. 
For instance, beyond the conventional reconstruction task, one could introduce an additional classification loss within the training objective.
This strategy encourages the latent representation to acquire useful features for classification tasks, which will subsequently be embodied in the synthesized data. 
\end{takeaway}


%% file: contents/4.1_adversarial.tex
\subsection{PPDS Based on Minimax Training}
\label{subsec:deep-at}

\subsubsection{Preliminaries} \quad
Minimax training involves training two components in opposition to each other in a zero-sum game.
A prominent example
in the context of generative modeling is the Generative Adversarial Network (GAN)~\cite{goodfellow2014generative}.
GANs consist of a generator network that produces samples resembling real data, and a discriminator network that distinguishes between the generated and real data. 

In contrast to other categories of generative modeling, 
a distinctive characteristic of GAN is that it consists of two interdependent models that are alternatively updated during the training procedure.
The escalation in system complexity warrants a reflection on the threat model, specifically, what information the adversary could access in order to launch privacy attacks.
This is critical in designing DP algorithms, as any information accessible to the adversary, and only that information, should be protected.
The strongest adversary 
can access the entire information flow (\eg, outputs, gradients) throughout training. A slightly weaker adversary can access only the generator; this is a reasonable assumption given that only the generator is typically released for public use. The weakest adversary sees only the synthetic data.

\subsubsection{Paper Summaries} \quad
We review two strategies of achieving DP, corresponding to the first two types of adversaries. The discussion of the model-agnostic weakest adversary is deferred to~\Cref{sec:future}. 

\noindent\textbf{Perturbing the Discriminator.}\quad
Existing literature primarily focuses on the strongest adversary.
We start with 
a preliminary method that is  implemented in several parallel studies
\cite{lu2017poster,beaulieu2019privacy,xie2018differentially,liu2019ppgan,zhang2018differentially},
which employs DP-SGD exclusively to the discriminator.
This method effectively ensures DP for the entire system at all times, as it makes all information flow private.
Importantly, given that only the discriminator interacts with the private data, the generator also satisfies DP once the discriminator does so, due to the post-processing property.
Additionally, confining DP-SGD to the discriminator circumvents introducing excessive noise to the generator, which could potentially impede the training process.

Extending GAN from the image domain to tabular data,
Frigerio \etal~\cite{frigerio2019differentially} adopt the one-hot representation for the categorical inputs and adjust the generator's output accordingly.
CTGAN~\cite{xu2019modeling} proposes a new normalization approach for the numerical inputs by modeling it as a variational Gaussian mixture model~(VGM)~\cite{bishop2006pattern}.
These adjustments in data representation and model architecture 
enable effective training of GAN and integration of DP-SGD~\cite{frigerio2019differentially,wang2020part,astolfi2021generating}.
Another line of work leverages the auto-encoding mechanism to automatically learn the representation for mixed-type tabular data, thereby circumventing the need for manual design of such representations.
A detailed discussion of these works is offered in~\Cref{subsec:deep-ae}.
\noindent\textbf{Perturbing the Discriminator's Signal.}\quad
In the case where only the generator is accessible to the adversary, 
it suffices to build a \textit{privacy barrier} between the discriminator and the generator.
Concretely, the training of the generator relies on the signal from the discriminator; perturbing the signal provides a privacy guarantee for the generator. 
Different works focus on different types of signal and different approaches of perturbation.

\textit{\underline{Partial gradient sanitization.}}\quad
GS-WGAN~\cite{chen2020gs} points out that the gradient of the generator can be decomposed into two components via the chain rule: 
an upstream gradient (discriminator loss w.r.t. the generated image) and a local gradient (generated image w.r.t. the generator's parameters). 
The authors perturb the upstream gradient via the Gaussian mechanism, as it is the sole bearer of private information. 
This method effectively sanitizes the generator's gradient without inducing substantial distortion, leading to more informative gradient updates and improved training outcomes.

\textit{\underline{Private knowledge aggregation and transfer.}}\quad
In the teacher-student framework~\cite{hinton2015distilling}, the knowledge of the teacher ensemble is aggregated and transferred to the student. 
In GANs, the discriminator acts as the teacher, providing signal (knowledge) to the generator (student). 
By privatizing this knowledge transfer, a privacy barrier is established between the discriminator and the generator.

Existing works consider two types of knowledge and employ different architectures 
for aggregation. 
PATE-GAN~\cite{yoon2018pategan} perceives the discriminator as a classifier and utilizes its output (true/fake) as knowledge. 
It extends PATE~\cite{papernot2017semi,papernot2018scalable}, a DP classification scheme, to privatize the discriminator's output. Specifically, they train multiple teacher discriminators along with a single student discriminator. The student learns from noisy labels that are obtained through privately aggregating the teachers' votes.
The update of the generator is then guided by the signal from the student discriminator.
In comparison, G-PATE~\cite{long2021gpate} and DataLens~\cite{wang2021datalens} focus on the discriminator's gradient as knowledge. 
They employ an architecture that comprises an ensemble of teacher discriminators and a student generator. 
The student generator receives private gradients that are obtained through the private aggregation of the teachers' gradients.
To mitigate the impact of noise,
G-PATE~\cite{long2021gpate} performs dimensionality reduction on the gradients using random projection, whereas
DataLens~\cite{wang2021datalens} conducts stochastic gradient compression in selective dimensions to reduce sensitivity.


\begin{takeaway}[Takeaways]
Researchers should tailor the development of PPDS algorithms according to the specific privacy requirements and threat models of each scenario. Applying existing privacy mechanisms indiscriminately to complex systems could lead to unwarranted reductions in data utility. 


\end{takeaway}
\begin{takeaway}[Open Problems]
Existing works commonly attain DP in minimax training by applying DP-SGD to conventional generative models. 
An open problem, however, lies in the design or identification of architectures that are better suited to DP-SGD.
Leveraging Neural Architecture Search (NAS)~\cite{elsken2019neural} could serve as a compelling approach, optimizing not only task performance but also compatibility with DP-SGD. This might involve a search strategy related to factors such as stable convergence, smooth gradient, and model sensitivity. 



\end{takeaway}

%% file: contents/4.5_optimal.tex
\subsection{PPDS Based on Optimal Transport}
\label{subsec:deep-ot}
\subsubsection{Preliminaries} \quad
Generative modeling based on optimal transport~\cite{genevay2018learning,salimans2018improving,patrini2020sinkhorn} adopt the Wasserstein distance~(W-distance)~\cite{villani2009wasserstein} to measure the discrepancy between probability distributions. 
Minimizing this metric facilitates the generation of samples that resemble a target distribution.
The Sinkhorn divergence~\cite{genevay2018learning} provides a smooth approximation of the W-distance by incorporating a regularization term, eliminating the need for using a discriminator to estimate the W-distance as in WGANs~\cite{arjovsky2017wgan}. 
This results in more efficient and stable training of a generative model, making it more robust to noise addition and suitable for DP.

\subsubsection{Paper Summary} \quad
DP-Sinkhorn~\cite{cao2021dpsinkhorn} minimizes the Sinkhorn divergence between the real and the generated distributions in a differentially private manner via partial gradient sanitization~\cite{chen2020gs} (see~\Cref{subsec:deep-at}).
To balance the bias-variance trade-off, the authors introduce a semi-debiased Sinkhorn loss. This effectively tackles the limitations of previous empirical Sinkhorn losses, which are either biased~\cite{feydy2019interpolating} or have large variance~\cite{salimans2018improving}. 
The semi-debiased loss is computed by partitioning the generated data and evaluating both the internal divergence (self-term loss) as well as their divergence with the real data (cross-term loss).
Privacy is ensured by sanitizing the gradient of the cross-term loss with respect to one partition of the data. 
The power of optimal transport and the Sinkhorn divergence makes DP-Sinkhorn excel at capturing multiple modes and converge stably.




\begin{takeaway}[Takeaways]
    Utilizing Sinkhorn divergence enables a tractable and efficient approximation of the W-distance. 
    In turn, this allows for the direct alignment of distributions, thereby eliminating the need for a discriminator and bypassing the instability issues inherent in minimax training. Moreover,
    the formulation of a distribution-aligning objective has the added benefit of mitigating mode collapse, a common pitfall in GANs.

\end{takeaway}


\begin{takeaway}[Open Problems]
    The cost function in optimal transport plays a crucial role in data generation. 
    Conventional choices involve using $L_1$ and $L_2$ distances, 
    which promote image-level similarity but may not align with the desired properties for PPDS.
    An open problem is thus to design task-specific cost functions, potentially using kernel-based methods to capture complex features and enhance data utility. This direction, which goes beyond simple pixel-level resemblance, could lead to a more effective balance between data utility, privacy preservation, and generation quality.
\end{takeaway}

%% file: contents/4.3_features.tex
\subsection{PPDS Based on Feature Alignment}
\label{subsec:deep-fa}

\subsubsection{Preliminaries} \quad
Generative models underpinned by feature alignment aim to learn a mapping from a noise distribution to the data distribution. 
This is achieved by aligning the features of the synthesized data with those of the real data, typically employing feature-level distances. 
This approach holds particular merit for PPDS as it ensures that the learned representation is useful for downstream tasks, without enforcing similarity in the data space. 

In feature alignment, 
DP is achieved by adding noise to the features of the real data distribution, a technique known as \textit{feature sanitization}. 
The generator built on
these sanitized features is DP due to the post-processing property. 

\subsubsection{Paper Summaries}\quad
{Kernel Mean Embeddings~(KMEs)}~\cite{smola2007hilbert} transform a probability distribution into a reproducing kernel Hilbert space~(RKHS)~\cite{berlinet2011reproducing} by computing the mean of kernel evaluations. 
Maximum Mean Discrepancy (MMD)~\cite{gretton2006kernel} quantifies the distance between two probability distributions within the RKHS, thus offering a mechanism to impose distributional constraints in generative modeling.
To achieve DP, one can add noise to the KME of the real data, then minimize the MMD between this noisy KME and the KME of the generated data~\cite{harder2021dp,vinaroz2022hermite,harder2022differentially}.

The computation of KMEs is resource-intensive. Fortunately, an efficient approximation of KMEs can be achieved by employing the inner product of feature vectors~\cite{scholkopf1998nonlinear}.
DP-MERF~\cite{harder2021dp} adopts the random Fourier features (RFFs)~\cite{rahimi2007random} to approximate KMEs.
However, to achieve good approximation quality, a large number of random features are required.
This translates into a high-dimensional feature vector,  resulting in a significant amount of noise.
DP-HP~\cite{vinaroz2022hermite} circumvents this challenge by using Hermite polynomial features~(HPFs)~\cite{hermite1864nouveau}, which encapsulate a higher degree of information within a smaller order of features compared to RFFs.
In the image domain,
perceptual features~(PFs), \ie, the concatenation of activations from all layers of a pre-trained deep convolutional neural network (CNN)~\cite{krizhevsky2017imagenet}, have been effectively employed for generative modeling~\cite{santos2019learning}. 
Building on PFs obtained from publicly pre-trained networks, DP-MEPF~\cite{harder2022differentially} generates samples of high visual quality on CIFAR~\cite{cifar10} and CelebA~\cite{liu2015faceattributes} datasets, while maintaining small $\veps$ values such as 1 and 0.2, respectively.

\begin{takeaway}[Takeaways]
Feature alignment offers several advantages:
\begin{itemize}[leftmargin=*]
\item Feature sanitization is a one-time process and eliminates the concern of accumulating privacy costs during subsequent training.
\item The objective functions are simple and can be computed and optimized efficiently. For instance, DP-MERF~\cite{harder2021dp} completes training within seconds. 
\end{itemize}
\end{takeaway}
\begin{takeaway}[Open Problems]
In the realm of feature alignment based data synthesis, a captivating research opportunity lies in the exploration of dynamic, data-dependent feature selection. Could we build a meta-learning framework that identifies optimal features for different datasets or applications, like selecting disease-specific features for medical image generation? By evolving feature selection from a static decision into a context-aware, learning-based process, we could push the boundaries of privacy-preserving data synthesis. 

\end{takeaway}
    

%% file: contents/4.4_stochastic.tex
\subsection{PPDS Based on Stochastic Simulation}
\label{subsec:deep-ss}

\subsubsection{Preliminaries}\quad
Diffusion Models (DMs) leverage stochastic simulations to represent complex, high-dimensional data distributions,
thereby displaying remarkable capabilities in image synthesis~\cite{song2019generative,song2020improved,song2021scorebased,ho2020denoising,rombach2022high}. 
A defining property of DMs is their iterative generation mechanism, starting with random noise and progressively refining it into a high-fidelity image that resembles the target distribution.




\subsubsection{Paper Summaries}\quad
Different types of DMs vary in their approach to 
the \textit{forward process}, in which data is gradually transformed into noise, and the \textit{reverse process}, where data is generated from noise through a series of  denoising steps.
We review two variants of DMs~\cite{yang2022diffusion} and discuss how existing works achieve DP within these models. 



\noindent
\textbf{Score-Based Generative Models~(SGMs)}~\cite{song2019generative,song2020improved} utilize a neural network to approximate the gradient of the log-density, known as the \textit{score}~\cite{hyvarinen2005estimation}, of a target distribution.
During the training phase, SGMs simulate the forward process by perturbing the data with Gaussian noise. A noise-conditioned score network (NCSN) is then trained to estimate the score function for the noisy data distributions. 
During the generation phase, SGMs employ score-based sampling approaches that leverage the score network, which guides the iterative refinement in the reverse direction.
Common score-based sampling approaches include Langevin Monte Carlo~\cite{grenander1994representations,song2019generative}.

DPGEN~\cite{chen2022dpgen} introduces a novel approach for generating DP synthesized data by learning DP scores. It begins with the creation of a sanitized dataset, consisting of Gaussian-perturbed samples and their respective private scores. Then, an NCSN is trained on this sanitized data. 
To privatize the scores, the authors leverage Randomized Response (RR)~\cite{warner1965randomized}, which probabilistically associates each perturbed sample with a nearby neighbor rather than its original non-perturbed source. 
An advantage of RR is that it incurs minimal error for $\veps>1$. 
Additionally, once the sanitized dataset is created, extending the training duration solely enhances the generation quality without introducing additional privacy costs. 
Putting together, DPGEN can synthesize perceptually realistic images of high dimensionality (CelebA~\cite{liu2015faceattributes} and LSUN~\cite{yu15lsun} of size $128\times 128 \times 3$) at $\veps=5$.



\noindent\textbf{Stochastic Differential Equations~(Score SDEs)}~\cite{song2021scorebased,karras2022elucidating} 
establish a framework where the forward diffusion process is governed by an SDE consisting of a drift term and a diffusion term. 
Any such diffusion process can be reversed by solving another reverse-time SDE~\cite{anderson1982reverse}, which relies on the score function. Once the score function at each step is learned, we can solve the reverse-time SDE with numerical methods, effectively transforming the noise into real data.



DPDM~\cite{dockhorn2022dpdm} achieves DP by applying DP-SGD to the denoiser (similar to a score network).
The authors note that the denoiser often has low capacity, and is thus amenable to DP-SGD since the magnitude of noise grows linearly with the number of model parameters.
They demonstrate state-of-the-art~(SOTA) performance on various benchmarks.


\begin{takeaway}[Takeaways]
The sequential nature of DMs grants them a distinct advantage in PPDS. Unlike other generative modeling approaches that opt for a single-pass synthesis, DMs utilize a sequence of gradual refinements, fostering a smoother and more controlled generation process. 
The modest progression allows DMs to seamlessly incorporate the distortions necessitated by privacy requirements,  minimizing the impact on the utility of the synthesized data.
\end{takeaway}
\begin{takeaway}[Open Problems]
A critical open problem is to develop a privacy accounting scheme tailored to the sequential nature of diffusion models.   
Specifically, exploring step-wise privacy loss accounting during model training stands as a promising direction.
Such an approach has the potential to deliver a more accurate distribution of the privacy budget. It recognizes that different steps might contribute unevenly to overall privacy leakage, allowing for more precise tracking and mitigation.

\end{takeaway}

%% file: contents/6_evaluation.tex
\input{tables/results}

\vspace{-2mm}
\section{Evaluation}\label{sec:eval}
\vspace{-2mm}
We benchmark DL-based methods on the task of private image data synthesis, 
using MNIST~\cite{lecun1998mnist} and FashionMNIST~\cite{xiao2017fashion}, two datasets commonly adopted in the literature. The results are presented in~\Cref{tab:results}. We make two additional comments here. First, PPDS significantly lags behind its non-private counterpart in the image domain. As pointed out by Chen~\etal~\cite{chen2022dpgen}, 
very few works can generate meaningful $32\times 32$ images under a reasonable privacy budget. Second, we narrow our focus to image data to distinguish our study from a prior work~\cite{tao2022benchmarking}, which set a benchmark for DP tabular data synthesis. We refer readers to the Appendix for general method selection criteria in real-world scenarios. 

\subsection{Experimental Setups}
\label{subsec:eval-setup}
\noindent\textbf{Inclusion \& Exclusion Criteria.}\quad
We aim to build a comprehensive benchmark covering all PPDS milestone approaches from the papers accepted at top ML \& security conferences.
We first narrowed down the approaches by available code repositories on GitHub.
We then made an initial attempt to run each approach, following the instructions in their README and applying the hyper-parameters reported in their evaluation settings.
We exclude approaches that either 1) fail to work; or 2) demonstrate significant discrepancy in our re-implementation, in order to prevent any misrepresentation of their method.
In the end, we obtain a list of six approaches, with three (DPGAN~\cite{xie2018differentially}, DP-CGAN~\cite{torkzadehmahani2019dp}, and GS-WGAN~\cite{chen2020gs}) under the category of minimax training (\Cref{subsec:deep-at}), one (DP-MERF~\cite{harder2021dp}) under feature alignment (\Cref{subsec:deep-fa}), one (DPGEN~\cite{chen2022dpgen}) under stochastic simulation (\Cref{subsec:deep-ss}), and one (DP-Sinkhorn~\cite{cao2021dpsinkhorn}) under optimal transport (\Cref{subsec:deep-ot}).


\noindent\textbf{Evaluation Scenarios.}\quad
We evaluate the performance of these approaches on two standard scenarios ($\veps=10$, $\veps=1$) and two challenging scenarios (stringent privacy---$\veps=0.2$, and smaller population---half training set size under $\veps=10$)
We also evaluate the non-private scenario using the non-private version of each approach. 
This is used as a baseline.


\noindent\textbf{Evaluation Metrics.}\quad
We synthesize a dataset of the same size as the training set.
We follow the literature~\cite{chen2020gs,harder2021dp,cao2021dpsinkhorn} to evaluate the fidelity and  utility of the synthesized data.
For fidelity, we measure the Frechet Inception Distance~(FID)~\cite{heusel2017gans} and the Inception Score~(IS)~\cite{li2017alice}.
For utility, we train classifiers on the synthesized dataset and test their performance on the real test data. 
We adopt 13 classifiers in all, including convolutional neural network~(CNN), multi-layer perceptron~(MLP), and 11 other scikit-learn~\cite{scikit-learn} classifiers.
We measure the performance of these classifiers using classification \textit{accuracy}.
Due to limited space, we report the averaged accuracy on the 11 scikit-learn classifiers, denoted as ``Avg'' in~\Cref{tab:results}.
For each combination of DDPS approach and evaluation scenario, we report the results averaged over five runs of dataset generation and classifier training following~\cite{chen2020gs,harder2021dp}.

We refer readers to our website for detailed information on the experimental setups (\eg, hyper-parameter tuning) and additional experimental results (\eg, runtime comparisons).


\subsection{Highlighted Conclusions}
\label{subsec:eval-conclusion}
\noindent\textbf{DP-MERF is the best practice.}\quad 
Among the evaluated approaches, DP-MERF~\cite{harder2021dp} stands out as the only one that consistently delivers high classification accuracy across all scenarios. It demonstrates minimal performance loss compared to its non-private counterpart, indicating its compatibility with DP. In contrast, DPGEN degrades significantly in the small epsilon regime. Additionally, DP-MERF exhibits remarkable computational efficiency. Thus, we highly recommend DP-MERF as an all-purpose approach. 

We note that the visual quality (measured by FID and IS) of DP-MERF is not the best; yet it effectively learns the  key features for classification. 
This trait is beneficial for classification, but caution is needed when the synthesized dataset is applied to other types of downstream tasks.

\noindent\textbf{The pessimistic conclusion in Stadler \etal~\cite{stadler2021synthetic} does not fully extend to image data.} \quad
Stadler \etal conclude that synthetic tabular data either has poor privacy or utility.
However, our experimental results suggest a different narrative.
Although the general trend across all approaches is that one has to sacrifice a significant amount of utility for strong privacy, two methods (DP-Sinkhorn and DP-MERF) yield notable performance at $\veps=1$, which is typically viewed as stringent privacy that meets practical requirements~\cite{lee2011much}. 
In particular, DP-MERF demonstrates the possibility of simultaneously achieving high utility and strong privacy in synthetic data, implying that such solutions exist—it's simply a matter of discovering them.

\noindent\textbf{Fidelity and utility are positively correlated.}\quad
We compute the Kendall rank correlation coefficient~\cite{kendall1948rank} between utility and fidelity measures for the above approaches, using an average derived from six pairs of measures (three for utility and two for fidelity). This is further averaged across five scenarios (non-private, two standard, and two challenging). The obtained value is 0.47 for MNIST and 0.65 for Fashion-MNIST, indicating a strong positive correlation between fidelity and utility in image data. 


We refer readers to our website for more conclusions.

\begin{takeaway}[Recommended Practices]
Across the course of our replication study, we have pinpointed a number of concerns that permeate various works in the field. These issues are elaborated upon on our website, leading us to propose several best practice recommendations:
\begin{itemize}[leftmargin=*]
    \item Report the runtime and discuss the computational efficiency.
    \item Acknowledge and report any limitations, such as the incapability of generating labels.
    \item Furnish rigorous privacy analysis both in the paper and the associated code; elucidate the factors or parameters influencing the privacy analysis.
    \item Be cautious about inadvertent privacy leakage in algorithmic design (\eg, utilizing a classifier trained on private data for label assignment) and implementation  (see~\cite{stadler2021synthetic}).
    \item Provide explicit descriptions of all algorithmic steps.
\end{itemize}
    
\end{takeaway}









%% file: tables/results.tex
{
\setlength{\tabcolsep}{3pt} 
\begin{table*}[tbp]
    \centering

    \caption{\small \textbf{Main results.} We evaluate six DL-based approaches on two datasets under five scenarios.
    We measure the utility (by classifier accuracy) and fidelity (by FID and IS) of the synthesized data. 
    ``Avg'' refers to the average accuracy over 11 scikit-learn classifiers.
    Each number is averaged over five runs of synthetic data  generation and classifier training.
    The best performing approach is presented in \textbf{bold}, while the second best performing one is shown in \textit{italics}.
    }
    \label{tab:results}

\vspace{-3mm}

\resizebox{\linewidth}{!}{%
        \begin{tabular}{lrrrrr|rrrrr|rrrrr|rrrrr|rrrrr
}
    \toprule
    \multirow{3}{*}{\small \textbf{MNIST}}
     & \multicolumn{5}{c|}{\multirow{2}{*}{\small Non-private}} 
     & \multicolumn{10}{c|}{\small Standard}
     & \multicolumn{10}{c}{\small Challenging} \\
     & & & & & 
     & \multicolumn{5}{c}{$\veps=10$}
     & \multicolumn{5}{c|}{$\veps=1$}
     & \multicolumn{5}{c}{$\veps=0.2$}
     & \multicolumn{5}{c}{Half ($\veps=10$)} \\
     \cmidrule(lr){2-6}
     \cmidrule(lr){7-11}
     \cmidrule(lr){12-16}
     \cmidrule(lr){17-21}
     \cmidrule(lr){22-26}
     & 
     MLP & CNN & Avg & FID$\downarrow$ & IS$\uparrow$ & MLP & CNN & Avg & FID$\downarrow$ & IS$\uparrow$ & MLP & CNN & Avg & FID$\downarrow$ & IS$\uparrow$ & MLP & CNN & Avg & FID$\downarrow$ & IS$\uparrow$ & MLP & CNN & Avg & FID$\downarrow$ & IS$\uparrow$  \\ \midrule
DPGAN & 85.5 & 88.9 & 52.0 & 63.1 & 8.6 & 74.8 & 74.7 & 54.7 & 268.8 & 2.0 & 40.0 & 58.6 & 27.0 & 387.1 & 1.0 & 15.1 & 8.9 & 12.3 & 420.2 & \textit{1.3} & 69.7 & 71.2 & 48.2 & 295.1 & 1.7\\
DP-CGAN & 84.6 & 91.6 & \textit{71.3} & \textit{39.7} & 9.4 & 60.5 & 64.8 & 51.7 & 174.2 & 4.2 & 9.3 & 25.3 & 15.5 & 281.8 & 1.5 & 9.2 & 12.9 & 10.2 & 292.0 & 1.0 & 62.5 & 65.0 & 49.8 & 248.1 & 2.3\\
GS-WGAN & 85.7 & 84.9 & 60.3 & 64.3 & \textbf{9.8} & 79.5 & 78.3 & 55.7 & 60.6 & \textbf{8.2} & 27.0 & 13.2 & 16.3 & 392.5 & 1.0 & 8.3 & 10.5 & 9.7 & 489.6 & 1.0 & 73.2 & 77.7 & 46.2 & \textit{59.2} & \textit{7.7}\\
DP-MERF & 83.4 & 87.0 & 67.0 & 103.7 & 5.5 & \textit{80.4} & \textit{83.2} & \textit{69.4} & 103.5 & 6.3 & \textit{81.8} & \textit{83.9} & \textbf{66.6} & \textbf{112.0} & \textbf{6.4} & \textit{75.6} & \textit{79.1} & \textbf{54.5} & \textbf{130.9} & \textbf{5.7} & \textit{80.0} & \textit{82.6} & \textit{65.2} & 106.8 & 6.4\\
DP-Sinkhorn & \textit{89.4} & \textit{92.3} & 66.1 & 52.3 & \textbf{9.8} & 77.0 & 79.6 & 52.5 & \textit{78.0} & 5.9 & 61.5 & 61.9 & 33.6 & 200.0 & \textit{3.3} & 56.6 & 48.8 & 34.8 & 208.9 & 1.1 & 75.2 & 78.2 & 44.8 & 87.7 & 5.2\\
DPGEN & \textbf{95.7} & \textbf{98.2} & \textbf{83.5} & \textbf{6.2} & 7.9 & \textbf{95.0} & \textbf{98.2} & \textbf{78.5} & \textbf{9.3} & \textit{6.7} & \textbf{92.3} & \textbf{97.6} & \textit{48.5} & \textit{125.5} & 1.3 & \textbf{79.8} & \textbf{96.1} & \textit{36.5} & \textit{183.1} & 1.0 & \textbf{96.0} & \textbf{98.4} & \textbf{84.1} & \textbf{5.5} & \textbf{8.0}\\
\bottomrule \\
\end{tabular}
}

\resizebox{\linewidth}{!}{%
        \begin{tabular}{lrrrrr|rrrrr|rrrrr|rrrrr|rrrrr
}
    \toprule
    \multirow{3}{*}{\makecell[l]{\small \textbf{Fashion-}\\\textbf{MNIST}}}
     & \multicolumn{5}{c|}{\multirow{2}{*}{\small Non-private}} 
     & \multicolumn{10}{c|}{\small Standard}
     & \multicolumn{10}{c}{\small Challenging} \\
     & & & & & 
     & \multicolumn{5}{c}{$\veps=10$}
     & \multicolumn{5}{c|}{$\veps=1$}
     & \multicolumn{5}{c}{$\veps=0.2$}
     & \multicolumn{5}{c}{Half ($\veps=10$)} \\
     \cmidrule(lr){2-6}
     \cmidrule(lr){7-11}
     \cmidrule(lr){12-16}
     \cmidrule(lr){17-21}
     \cmidrule(lr){22-26}
     & 
     MLP & CNN & Avg & FID$\downarrow$ & IS$\uparrow$ & MLP & CNN & Avg & FID$\downarrow$ & IS$\uparrow$ & MLP & CNN & Avg & FID$\downarrow$ & IS$\uparrow$ & MLP & CNN & Avg & FID$\downarrow$ & IS$\uparrow$ & MLP & CNN & Avg & FID$\downarrow$ & IS$\uparrow$  \\ \midrule
DPGAN & \textit{79.3} & \textit{78.7} & 60.2 & \textit{64.0} & \textit{7.7} & 63.9 & 59.5 & 50.1 & 290.8 & 3.7 & 44.8 & 54.1 & 31.2 & 433.7 & 1.4 & 15.7 & 20.6 & 12.8 & 438.2 & 1.4 & 61.9 & 58.7 & 47.0 & 316.0 & 2.3\\
DP-CGAN & 64.3 & 64.5 & 56.6 & 147.6 & 5.2 & 53.5 & 49.0 & 37.9 & 256.5 & 3.3 & 24.5 & 18.0 & 21.0 & 339.1 & 1.8 & 12.0 & 12.8 & 11.2 & 331.2 & 1.7 & 36.4 & 37.8 & 35.4 & 351.3 & 1.3\\
GS-WGAN & 76.4 & 71.1 & 58.7 & 83.7 & \textbf{8.0} & 65.9 & 64.7 & 50.7 & 125.9 & \textit{5.7} & 21.6 & 12.5 & 19.8 & 334.1 & 1.0 & 9.7 & 16.0 & 9.4 & 490.5 & 1.0 & 67.8 & 65.2 & 53.4 & 118.3 & \textit{5.6}\\
DP-MERF & 74.8 & 73.0 & \textit{61.0} & 101.0 & 5.4 & \textit{73.5} & \textit{71.7} & \textit{58.2} & \textit{102.3} & 5.5 & \textbf{72.6} & \textbf{72.4} & \textbf{58.9} & \textbf{97.1} & \textbf{5.3} & \textbf{70.1} & \textbf{69.7} & \textbf{51.0} & \textbf{139.8} & \textbf{4.2} & \textit{73.6} & \textit{73.7} & \textit{56.9} & \textit{105.8} & 5.1\\
DP-Sinkhorn & 77.6 & 77.5 & 59.1 & 90.8 & 7.5 & 69.4 & 65.4 & 50.2 & 164.3 & 4.5 & \textit{58.0} & \textit{56.3} & \textit{38.5} & 216.2 & \textit{2.9} & \textit{45.9} & \textit{44.3} & \textit{30.7} & 260.5 & \textit{2.6} & 70.2 & 67.8 & 48.6 & 164.0 & 4.9\\
DPGEN & \textbf{80.2} & \textbf{84.5} & \textbf{71.2} & \textbf{14.4} & 7.3 & \textbf{79.5} & \textbf{84.0} & \textbf{70.2} & \textbf{11.0} & \textbf{7.3} & 36.3 & 41.7 & 21.3 & \textit{113.2} & 2.0 & 27.1 & 33.2 & 17.3 & \textit{190.0} & 1.2 & \textbf{80.2} & \textbf{84.6} & \textbf{70.6} & \textbf{12.2} & \textbf{7.3}\\
\bottomrule \\
\end{tabular}
}

\end{table*}
}

%% file: contents/7_future.tex
\vspace{-2mm}
\section{Future Directions}
\vspace{-2mm}
\label{sec:future}



\noindent \textbf{Emerging Applications.} \quad 
Current PPDS algorithms remain  inadequate for addressing the challenges posed by the high-dimensional unstructured data, which are prevalent in contemporary everyday applications. Such data types are exemplified by high-resolution images~\cite{bioimages,smith2022realistic}, audio files~\cite{donahue2018adversarial}, videos~\cite{wang2018video}, and text~\cite{bo2019er}. To mitigate this issue, it would be beneficial for researchers to leverage the most recent advancements in diffusion models~\cite{rombach2022high} and large language models (LLMs)~\cite{li2022large}, integrating them with privacy mechanisms to produce a more robust solution.





\noindent\textbf{Privacy Attacks.}\quad
Conventional membership inference attacks primarily operate on model predictions or low-dimensional feature vectors such as logits~\cite{shokri2017membership}. However, PPDS systems yield high-dimensional synthetic records, inherently richer in information. This poses the question of whether such increased information enables novel attacks on PPDS systems, thereby facilitating record linkage. Stadler~\etal~\cite{stadler2021synthetic} initiated attempts in this direction with synthetic tabular data. We advocate for further development of innovative attacks that can be applied to image data.


\noindent \textbf{Privacy Auditing.} \quad
Given a synthetic dataset released by a PPDS system, we would like to quantitatively assess its adherence to the claimed privacy guarantees. Previous work has demonstrated the feasibility of achieving such goal for a given private model through data poisoning~\cite{nasr2021adversary}. It warrants exploration as to whether comparable techniques can be effectively applied to PPDS systems.

\noindent \textbf{Privacy Notions.}\quad
When applying PPDS to pragmatic domains such as healthcare and IoT, DP may fall short due to its lack of intuitive interpretation.
In contrast, the level of privacy with respect to downstream tasks is often more interpretable and consequently frequently adopted. 
For context-aware tasks of this nature, the enforcement of DP may lead to unnecessary utility degradation. 
Therefore, it is crucial to develop privacy notions that are tailored for specific applications or tasks.


\noindent\textbf{Threat Models.}\quad
Existing literature has demonstrated that making assumptions about an adversary's knowledge could permit the use of a larger $\veps$, thereby enhancing utility~\cite{bhowmick2018protection}. In the context of PPDS, 
it is reasonable to formulate several plausible assumptions about the adversary, such as 
lack of access to the training process or even the trained model, with only the released data being available.
These assumptions could pave the way for a refined privacy analysis.

\noindent \textbf{Latent Issues of PPDS.} \quad
PPDS has been demonstrated to degrade fairness, leading to disproportionate impacts on underrepresented groups~\cite{cheng2021can,ganev2022robin}. 
Furthermore, the ranking of feature importance within the synthetic data is not sufficiently preserved~\cite{giles2022faking}. There are likely additional latent issues with PPDS that require thorough investigation.

\looseness=-1
Finally, we wish to underscore a dilemma pervading the research community: namely, \textit{the lack of a consensus on the selection of appropriate utility metrics and privacy notions for the evaluation of synthetic data}. Privacy notions, even when restricted to DP, are frequently misapplied; a scant number of studies incorporate privacy tests. Each paper appears to employ its unique suite of utility metrics, thereby risking an evaluation bias in favor of their specific algorithm. Moreover, there is a conspicuous dearth of task-centric utility metrics for trajectory and graph data. Consequently, we urgently advocate for the establishment of a comprehensive benchmark, replete with a consistent and exhaustive array of utility metrics and privacy tests. 

\vspace{1em}
\looseness=-1
\noindent\textbf{Acknowledgments.}\quad
We are extremely grateful to Varun Chandrasekaran for his help in improving this work.

%% file: tables/tab_ref_2.tex
\begin{table*}[htb]
\caption{\textbf{A reference table for the candidate algorithms of PPDS divided by data and task. } SM=statistical methods, DM=DL-based methods.  }
\label{tab:model_summary}
\centering
\setlength{\tabcolsep}{3pt}
\resizebox{\linewidth}{!}{
\begin{tabular}{llllllll}
\toprule
\textbf{Data Type}                       & \textbf{Applicable Task}                                                                                                    & \textbf{Method}               & \textbf{Approaches} &\multicolumn{2}{c}{\textbf{Most Cited}}&\multicolumn{2}{c}{\textbf{Most Recent}} \\ \cmidrule(lr){5-6}\cmidrule(lr){7-8}
&&&&Reference&\makecell[l]{Methodology/ \\Principle}&Reference&\makecell[l]{Methodology/ \\Principle} \\\midrule
\multirow{8}{*}{Tabular data}   & \multirow{2}{*}{Task-independent}                                                                      & SM                  &   
\makecell[l]{\citep{bindschaedler2017plausible,chen2015differentially,mckenna2021winning,cai2021data, zhang2021privsyn,zhang2017privbayes,li2014differentially}\\\citep{jiang2013differential,chanyaswad2019ron,asghar2019differentially,gambs2021growing, li2011compressive}       }
&PrivBayes~\citep{zhang2017privbayes} &Marginal-based  & PrivMRF~\citep{cai2021data}  &Marginal-based  \\ \cmidrule{3-8} 
                            &          & \multirow{1}{*}{DM}    &\makecell[l]{\citep{lu2017poster,xie2018differentially,liu2019ppgan,frigerio2019differentially,astolfi2021generating}\\
                            \citep{wang2020part,yoon2018pategan,long2021gpate}\\
                            \citep{acs2018differentially,chen2018differentially,tantipongpipat2021differentially,abay2018privacy}\\
                            \citep{harder2021dp,vinaroz2022hermite}
                            } & \multirow{1}{*}{PATE-GAN~\citep{yoon2018pategan}} & \makecell[l]{Minimax training}
                                        & DP-MERF~\cite{harder2021dp} &Feature alignment  \\    \cmidrule{2-8} 
                            & Statistical queries                                                                                                   & SM                  &\citep{hardt2012simple,vietri2020new,gaboardi2014dual,aydore2021differentially}
                            & MWEM~\citep{hardt2012simple} &Query-based &RAP~\citep{aydore2021differentially}  &Query-based    \\ \cmidrule{2-8} 
                            & Clustering/nearest neighbors                                                                                                & SM                  &\citep{kenthapadi2012privacy,xu2017dppro}       &PrivateProjection~\citep{kenthapadi2012privacy}   &Projection-based  &Dppro~\citep{xu2017dppro} &Projection-based   \\ \cmidrule{2-8} 
                             & Network traffic synthesis                                                                                      &DM                     & \citep{fan2021dpnet}       &DPNeT~\citep{fan2021dpnet}    &Minimax training &DPNeT~\citep{fan2021dpnet} &Minimax training  \\ \cmidrule{2-8}
                            & Health record synthesis                                                                                     &DM                     & \citep{lee2020generating,torfi2022differentially,beaulieu2019privacy} &RDP-CGAN~\citep{torfi2022differentially}    &Minimax training       &RDP-CGAN~\citep{torfi2022differentially} &Minimax training  \\ \hline
\multirow{1}{*}{Trajectory data} & \multirow{1}{*}{Task-independent}                                                                                        & SM &  \citep{chen2012differentially,chen2012ngrams,he2015dpt,zhang2016privtree,mir2013dp,roy2016practical,gursoy2018differentially,gursoy2018utility,gursoy2020utility, wang2017protecting}      &PrefixTree~\citep{chen2012differentially}   &Tree-based &OptaTrace~\citep{gursoy2020utility}    &Distribution-based   \\  \hline
{Graph data}      & {Task-independent}                                                                     & SM                  &\citep{mir2012differentially,sala2011sharing,wang2013preserving,lu2014exponential,xiao2014differentially,jorgensen2016publishing,chen2020publishing}        &dK~\citep{sala2011sharing}    &dK-Graph &C-AGM~\citep{chen2020publishing}  &AGM   \\  \hline 
{Image data}      & {Task-independent}                                                                  & {DM}                & \makecell[l]{\citep{torkzadehmahani2019dp,zhang2018differentially,liu2019ppgan,chen2020gs}\\
\citep{chen2022dpgen,chen2018differentially,harder2021dp,cao2021dpsinkhorn,vinaroz2022hermite,harder2022differentially,dockhorn2022dpdm,long2021gpate,wang2021datalens}}
& {DPGAN~\citep{xie2018differentially}}  & {Minimax training}  & {DPDM~\citep{dockhorn2022dpdm}} & {\makecell[l]{Stochastic simulation}}     \\  

\bottomrule
\end{tabular}
}
\end{table*}